\documentclass[extra, mreferee]{gji}
\usepackage{timet,color}
\usepackage[urlcolor=blue,citecolor=black,linkcolor=black]{hyperref}
\usepackage{amsmath}
\usepackage{amsfonts}
\usepackage{graphicx}

\title[Monogenic signal analysis for wavefield characterisation]
  {Monogenic signal analysis for wavefield characterisation in distributed acoustic sensing}
\author[A. Seguí, A. Ugalde, A. Fichtner, H. Van Luong and W. Armour ]
  {Aleix Seguí$^{1}$, Arantza Ugalde$^{2}$, Andreas Fichtner$^{3}$, \and Huynh Van Luong$^{4}$ and Wes Armour$^{1}$ \\
   $^{1}$ Department of Engineering Science, University of Oxford, Oxford, United Kingdom \\
   $^{2}$ Institute of Marine Sciences, CSIC, Barcelona, Spain \\
   $^{3}$ Department of Earth and Planetary Sciences, ETH Zürich, Zürich, Switzerland \\
   $^{4}$ AP Sensing GmbH, Böblingen, Germany \\
   }

\date{Received 1998 December 18; in original form 1998 November 22}
\pagerange{\pageref{firstpage}--\pageref{lastpage}}
\volume{200}
\pubyear{{\the\year{}}}

\begin{document}

\label{firstpage}

\maketitle

\begin{summary}
The spatio-temporal wavefield sampled by distributed acoustic sensing (DAS) contains intertwined information about signal strength, oscillatory structure and propagation. We introduce a methodology to disentangle these complementary properties, with applications to the analysis of weak arrivals obscured by noise or stronger wave trains. The method is based on the monogenic signal, a two-dimensional extension of the analytic signal defined via the Riesz transform. In DAS, the monogenic signal decomposes locally dominant arrivals into their amplitude, phase and apparent slowness along the fibre. Moreover, the consistency of the wavefield orientation, quantified using the notion of directional coherency,  allows coherent arrivals to be identified and enhanced. In this work, we apply the monogenic signal to analyse submarine \textit{T}-wave recordings following the magnitude 3.5 earthquake of 5 October 2023 in the western Alboran Sea, acquired along approximately 60 km of the Estepona--Tetouan fibre-optic cable. Directional coherency allows us to enhance coherent wave trains, strongly attenuating the incoherent background while preserving their oscillatory structure. Combining the monogenic attributes, we track individual oscillations across the fibre, obtaining spatially consistent apparent-slowness estimates. Together with varying cable geometry, these estimates allow us to infer the incoming directions of two wave trains, revealing distinct backazimuths. Furthermore, we extend the methodology to direction-selective enhancement, revealing a weaker third wave train with opposite moveout that is largely obscured by stronger arrivals. Beyond these examples, we show that monogenic analysis offers new possibilites for wavefield comparison, with promising applications to lossy data compression and inversion. Together, our results highlight the versatility of monogenic analysis and its potential to become a fundamental part of the toolkit for interpreting DAS recordings.
\end{summary}


\section{Introduction}

The amplitude envelope and instantaneous phase provide complementary descriptions of an oscillatory signal and are well-established attributes for interpreting seismic traces. Both arise naturally from the analytic signal, in which a trace and its Hilbert transform form the real and imaginary components of a complex-valued representation. Expressing the signal in polar form separates the magnitude of the oscillation from its position within the oscillatory cycle, allowing its energetic and structural information to be examined independently. For two-dimensional signals, this construction is generalised by the monogenic signal, introduced by \citet{felsberg2001monogenic}, which replaces the Hilbert transform with the vector-valued Riesz transform. The key feature of the monogenic signal is that it supplements local amplitude and phase with a local orientation attribute that describes the predominant direction of oscillation.

The monogenic signal was originally motivated by the need to demodulate different aspects of spatially varying patterns, with applications including local orientation estimation, edge detection, image analysis, and denoising \citep{felsberg2001monogenic}. Since its introduction, this decomposition has supported applications across a range of fields. In cardiac imaging, for example, \citet{alessandrini2013myocardial} used monogenic phase to track deformation of the heart muscle independently of variations in image intensity and contrast between frames. The amplitude envelope, by contrast, describes variations in the strength of a pattern independently of its oscillations. \citet{mateling2022turbulentflow} used this information to investigate how the amplitudes of structures in different regions of a turbulent flow vary together. The complementary roles of amplitude and phase are also evident in guided-wave imaging, where envelopes allow the strengths of different wavelength components to be compared \citep{flynn2013structuralimaging}, while spatial variations in phase provide estimates of local wavelength following mode filtering \citep{spytek2021meanlocalfrequency}.

In geophysical applications, monogenic attributes have been used to characterise structures with contrasting amplitudes and geometries. \citet{hidalgo2015edgedetectionpotentialfield} used the monogenic phase to delineate magnetic lineaments associated with strong and weak anomalies, and \citet{picard2018seafloordescription} applied monogenic analysis to describe the homogeneity, anisotropy, and complexity of seafloor textures in sonar images. In seismic reflection profiles, phase provides information about the arrangement and continuity of reflectors, whereas the amplitude envelope retains information about their strength. \citet{xhao2020horizonpicking} used these complementary descriptions for horizon picking in subbottom profiles, and \citet{wang2019seismicgeologicalstructure} proposed monogenic phase as a measure of coherence for describing geological structure from seismic traces.

In the present work, we consider the dense spatio-temporal sampling of the seismic wavefield provided by distributed acoustic sensing (DAS). Over the past decade, DAS has become a widely used tool for large-scale monitoring across geophysics, seismology and engineering \citep[e.g.,][]{Hartog_2017,Lindsey_2021}. DAS has enabled observations over spatial scales and in environments that are difficult to instrument conventionally, including boreholes \citep[e.g.,][]{Mateeva_2014,Daley_2016, Brisbourne_2021}, urban areas \citep[e.g.,][]{Lindsey_2017,Biondi_2017,Spica_2020,Bowden_2026}, the ocean floor \citep[e.g.,][]{Spica_2020b,Lior_2021,ugalde2021noiselevels}, active volcanic systems \citep[e.g.,][]{Klaasen_2021,Jousset_2022,Klaasen_2022}, and the cryosphere  \citep[e.g.,][]{Booth_2020, Hudson_2021, Fichtner_2025}.

Unlike conventional seismic sensors,  which measure particle motion at discrete locations, DAS channels record axial strain or strain rate averaged over a finite gauge length, such that the recorded response depends on the spatial wavelength of the wavefield and on the projection of the associated deformation onto the fibre direction \citep{dean2017gaugelength,kennett2022seismicwavefield}. Therefore, DAS measurements represent a directional and spatially averaged response of the seismic wavefield. More fundamentally, \citet{muir2022wavefieldbasedevaluation} advocated treating a DAS array as a sampling of an underlying spatial wavefield rather than as a collection of individual channels. 

Building on this view, we propose the monogenic signal as a characterisation technique of DAS recordings. We apply the analysis jointly to distance along the fibre and time coordinates, treating the data as two-dimensional oscillatory fields. In regions where the wavefield is locally dominated by a coherent arrival forming a ridge, the monogenic envelope and phase describe the local amplitude and phase of the arrival, while the monogenic orientation describes the direction normal to the ridge in distance--time space. The ridge direction, obtained from this orientation, defines the local slope and directly relates to the apparent slowness of the arrival along the fibre. 

Thus, our main proposition is that the monogenic representation provides, within a common local decomposition, the amplitude, phase and apparent propagation slowness of a dominant coherent arrival. Together, these attributes constitute a compact characterisation of the sensed wavefield.

Among these attributes, apparent slowness is particularly important in seismic analyses because it characterises the local moveout of coherent wavefield features along the fibre. It is used to separate wavefield components, analyse surface-wave dispersion and velocity structure, and characterise sources through array-processing methods, and it also enters directly into the conversion between strain and ground-motion quantities. For a locally plane wave, strain and particle velocity are related through the apparent slowness along the fibre \citep{wang2018groundmotion}. Several approaches therefore estimate propagation velocity from neighbouring DAS channels, such as frequency--wavenumber analysis \citep{wang2018groundmotion}, local slant stack \citep{lior2021straintogroundmotion} and beamforming \citep{nasholm2022arraysignalprocessing}. Complementary to these methods, our monogenic attribute instead originates from the local field orientation and jointly supplies the amplitude and phase of the locally dominant arrival, distinguishing changes in recorded strength from changes in local moveout. This separation is relevant because variations in recorded amplitude need not correspond to changes in propagation \citep{muir2022wavefieldbasedevaluation}, with fibre--medium coupling further complicating amplitude interpretation \citep{hudson2025unlockingamplitude}. 

In addition, to obtain robust estimates of the monogenic attributes, we introduce a notion of a joint local-neighbourhood. We combine the local Riesz components through a spatio--temporally averaged tensor, such that consistently oriented contributions reinforce a common direction while fluctuations with inconsistent orientations are attenuated by the averaging. This method provides a stabilised monogenic estimates, while also allowing to quantify the degree to which a single Riesz orientation--or slowness-- dominates the neighbourhood \citep{unser2009multiresolution}. We refer to this latter quantity as directional coherency, and use it to identify and enhance coherent arrivals. Specifically, directional coherency enables the disambiguation of arrivals obscured in a stochastic and incoherent background, with tensor-averaged estimates being robust for physical interpretation. 

Beyond the theoretical and methodological contributions, we empirically test the monogenic framework on submarine DAS recordings of oceanic \textit{T}-wave trains acquired in the western Alboran Sea. These multi-channel wavefields demonstrate how the four derived attributes--amplitude, phase, slowness, and directional coherency--operate synergistically across different stages of wavefield interpretation. First, we use directional coherency to suppress background noise and enhance coherent wave trains. Next, we combine monogenic amplitude and directional phase to isolate individual oscillatory cycles, while the orientation-derived apparent slowness guides cycle association across adjacent channels. We then show that these associated slowness estimates, projected along curved cable segments, provide robust constraints on the horizontal propagation backazimuth of the arriving wave packets. Finally, we explore the sensitivity of the monogenic representation to controlled wavefield perturbations and discuss its utility for pairwise comparison of fields, with applications to simulation, lossy data compression, and inversion misfits. The proposed methods are made available through RieszLab, an open-source software package developed to facilitate their broader adoption within the research community.

\section{Methodology}

\subsection{Theoretical background}

We begin our development with a review of the Hilbert transform for one-dimensional traces which is used to define the analytic signal. We then turn to the Riesz transform, which is a natural extension of the formalism to two dimensions, and use it to define the monogenic signal. We discuss the interpretation of the transforms and their components. 

\subsubsection{Hilbert transform}

The Hilbert transform of a real-valued function \(f(t)\) is a singular integral operator defined by convolution with the kernel \(h(t)=1/(\pi t)\) as
\begin{equation}
    \mathcal{H}\{f\}(t)
    =
    \frac{1}{\pi}
    \lim_{\varepsilon \to 0^{+}}
    \int_{|t-\tau|>\varepsilon}
    \frac{f(\tau)}{t-\tau}\,d\tau.
\end{equation}
The integral is interpreted in the Cauchy principal-value sense because the kernel is singular at the origin \(\tau=t\) \citep[see, e.g., ][]{Grafakos2014}. The transform is well-defined for finite-energy signals in the mean-square sense, and is both linear and translation invariant. Letting $F(\omega)=\mathcal{F}\{f\}(\omega)$
denote the Fourier transform of \(f(t)\), the Hilbert transform has the frequency-domain representation
\begin{equation}
    \mathcal{F}\{\mathcal{H}\{f\}\}(\omega)
    =
    -i\,\operatorname{sign}(\omega)\,F(\omega),
\end{equation}
where $i$ is the imaginary unit. This expression reveals the Hilbert transform as an operator that shifts the spectral components of a signal by $90^\circ$ in phase, so that cosine functions are mapped to their corresponding sine functions. The sign term is needed to preserve Hermitian symmetry, ensuring that the signal stays real-valued.

The Hilbert transform is a central piece in the definition of the complex analytic signal. For a real-valued function $f(t)$, its analytic signal is defined as
\begin{equation}
f_a(t)
=
f(t)+i\mathcal{H}\{f\}(t)
=
a(t)e^{i\phi(t)},
\end{equation}
where $a(t)=|f_a(t)|$ and $\phi(t)=\arg f_a(t)$ are the instantaneous amplitude and phase, respectively. For example, for the elementary
case $f(t)=A\cos(\omega_0t)$, where $\omega_0$ is an angular frequency, we find 
$\mathcal{H}\{f\}(t)=A\sin(\omega_0t)$ and therefore
$f_a(t)=Ae^{i\omega_0t}$. The modulus then recovers the amplitude of the oscillation, while the argument specifies its position within the
oscillatory cycle.  These quantities are widely used in geophysics and seismology, where instantaneous amplitude, phase, and frequency form a standard set of attributes for characterising variations in the recorded wavefield \citep[e.g., ][]{Taner1979,Barnes2007}. 

\subsubsection{Riesz transform}

The Riesz transform provides a related construction for signals of more than one dimension. It is the natural multidimensional analogue to the Hilbert transform in the sense that it is defined with the kernel $\mathbf{h}(\mathbf{x}) = c_d \, \mathbf{x} / |\mathbf{x}|^{d+1},$ applied to $d$-dimensional vectors $\mathbf{x}$, with the normalisation constant $c_d = \Gamma((d+1)/2)/\pi^{(d+1)/2}$ ensuring that the combined Riesz-transform components preserve the $L^2$ energy of the original signal \citep[e.g., ][]{Grafakos2014}. In two dimensions, the Riesz transform of a function $f(\mathbf{x})$ with $\mathbf{x} = (x_1, x_2)$  has two components $j=1,2$ that are defined as
\begin{equation}
    \mathcal{R}_j\{f\}(\mathbf{x})
    =
    \frac{1}{2\pi}
    \lim_{\varepsilon \to 0^{+}}
    \int_{|\mathbf{x}-\mathbf{y}|>\varepsilon}
    \frac{x_j-y_j}{|\mathbf{x}-\mathbf{y}|^{3}}
    f(\mathbf{y})\,d\mathbf{y},
\end{equation}
and as in the earlier case, is interpreted as a principal value integral. It is an isometrical scalar-to-vector valued transform that commutes with translation and scaling, and is rotation-equivariant, such that dilation of the input produces the corresponding dilation of the transformed field, and if a signal is rotated, the two Riesz components rotate as a vector with the signal. For additional detail, we provide an explanation and proof of these properties in Appendix~\ref{sec:properties_riesz}.

The frequency response of each component is
\begin{equation}
\mathcal{F}\{\mathcal{R}_j f\}(\boldsymbol{\omega})
=
-i\frac{\omega_j}{\|\boldsymbol{\omega}\|}
F(\boldsymbol{\omega}),
\end{equation}
with the multipliers of both components ($j=1,2$) defined to be zero at $\boldsymbol{\omega}=\mathbf{0}$. Thus, the Riesz transform only modifies the angular part of the Fourier spectrum, as is the case of the Hilbert transform. To illustrate its behaviour, consider a simple plane wave $f(\mathbf{x})
=
A\cos(\boldsymbol{k}\cdot\mathbf{x})$ with $\boldsymbol{k} = (k_1, k_2)$ being its wavenumbers. Its Riesz components are $\mathcal{R}_j \{ f(\mathbf{x}) \}
=
A\, (k_j /\|\boldsymbol{k}\|)
\sin(\mathbf{k}\cdot\mathbf{x}),$ which means that they contain the same $90^\circ$ phase-shifted oscillation, weighted by the
components of its propagation direction. In fact, the two Riesz components can be combined to construct an analogue of the analytic signal along any chosen direction. Specifically, for a unit vector $(\cos\theta,\sin\theta)$ with orientation $\theta$, the linear combination
\begin{equation} \label{eq:dir_hilb}
    \mathcal{H}_{\theta} \{f\} = \cos\theta\,\mathcal{R}_1 \{f\} + \sin\theta\,\mathcal{R}_2 \{f\}
\end{equation}
defines a directional Hilbert transform \citep{unser2009multiresolution}. For a locally unidirectional signal, when the orientation coincides with its local direction of variation,  the complex-valued signal $f+i\mathcal{H}_{\theta}\{f\}$ is interpreted as the analytic signal obtained through the regular Hilbert transform along the variation axis.

Another interesting view is that the Riesz transform behaves similarly to the gradient: the two operators have the same directional dependence in the Fourier domain, but the Riesz transform normalises the gradient magnitude by radial frequency. As such, it provides a natural measure of directionality without the preferential amplification of high frequencies associated with differentiation \citep{kothe2005}. For this reason, the Riesz transform is convenient for constructing derivative-like operators, including directional and higher-order derivatives and has applications in the study of partial differential equations \citep{Grafakos2014}.

\subsubsection{Monogenic signal}

The constructions so far provide the basis for a two-dimensional analogue of the analytic signal. Following \citet{felsberg2001monogenic}, the
monogenic signal of a real-valued function $f(\mathbf{x})$ is the three-component field
\begin{equation}
\mathbf{f}_m(\mathbf{x})
=
\left(
f(\mathbf{x}),
\mathcal{R}_1 \{ f \}(\mathbf{x}),
\mathcal{R}_2 \{ f\}(\mathbf{x}) 
\right).
\end{equation}
Letting $r_j = \mathcal{R}_j \{ f \}$, the vector modulus defines the local amplitude $A(\mathbf{x}) = \sqrt{f^2 + r_1^2 + r_2^2 }$, with the local phase $\phi(\mathbf{x})$ and local orientation $\theta(\mathbf{x})$ satisfying the spherical coordinate relation
\begin{equation}
    f = A \cos \phi, \  r_1 = A \sin \phi \cos\theta, \   r_2 = A \sin \phi \sin \theta,
\end{equation}
mirroring the polar form of the one-dimensional analytic signal. 

\begin{figure*}
    \centering
    \includegraphics[width=0.9\linewidth]{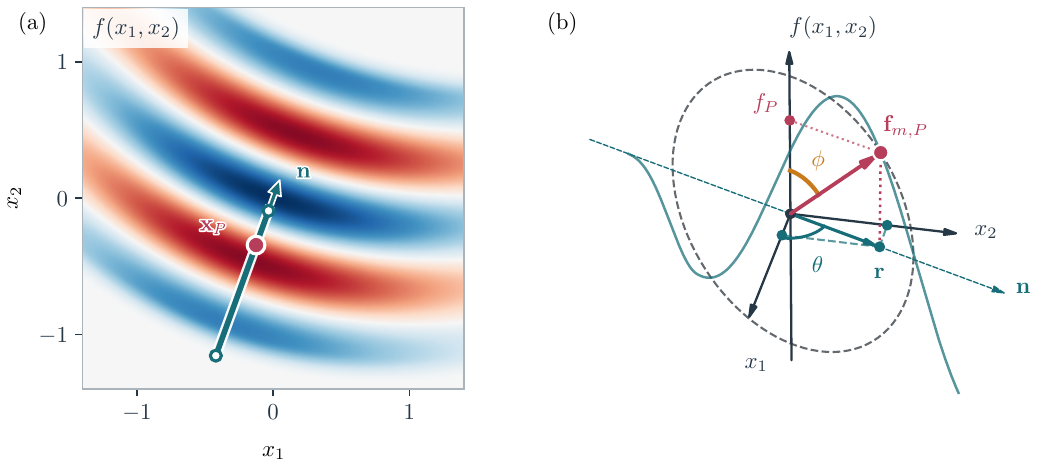}
    \caption{Illustration of the monogenic signal computed for an oscillatory field. 
    (a) An oscillatory field on dimensions $(x_1,x_2)$ with gently curved wavefronts. The highlighted segment follows the local normal $\mathbf n$ through the selected point $ \mathbf x_P$, defining a profile along the normal. 
    (b) The monogenic signal vector $\mathbf f_{m,P}$ combines the field value $f(\mathbf x_P)$, plotted vertically, with the Riesz vector $\mathbf r=(r_1,r_2)^{\mathsf T}$ in the horizontal plane. The angle $\theta$ describes the orientation of the normal $\mathbf n$, with the profile highlighted in (a) being overlaid in the vertical plane containing this direction. Within this plane, the monogenic vector is represented by the directional analytic signal $f(\mathbf x_P)+\mathrm i\,\mathcal H_{\theta} \{ f(\mathbf x_P) \}$. This reveals how the polar angle $\phi$ corresponds to an instantaneous phase.   }
    \label{fig:monogenic}
\end{figure*}

\citet{unser2009multiresolution} provided an insightful way to understand these quantities through the directional Hilbert operator (Equation \ref{eq:dir_hilb}). First, notice that $\mathcal{H}_\theta \{f\}$ acts as the dot product between the Riesz vector $(r_1, r_2)$ and a directional unit vector. Because the monogenic orientation $\theta$ indicates exactly the Riesz vector direction, $\theta$ then corresponds to the direction that maximises that dot product, giving $|\mathcal{H}_{\theta} f| = \sqrt{r_1^2 + r_2^2}$. Once oriented along this predominant direction, the two-dimensional problem effectively reduces to one dimension: we can construct the directional analytic signal $f + i \mathcal{H}_{\theta} \{ f \}$, whose standard instantaneous amplitude and phase are identical to the monogenic amplitude $A$ and phase $\phi$. We describe this geometric relationship in Figure~\ref{fig:monogenic}, where we identify the monogenic attributes for a point in a localised wavefield.

Consequently, the monogenic amplitude describes the local strength of the oscillation, the phase describes the position within the oscillatory cycle,
and the orientation describes its local geometric direction. Its physical interpretation depends on the physical quantities being measured. 

\subsection{Monogenic signal in DAS}
\label{sec:monogenic_das}

We now present our formulation of the monogenic signal applied to the spatio-temporal field recorded by DAS. Following \citet{kennett2022seismicwavefield}, consider a monochromatic plane wave propagating through three-dimensional space. Let the wave have displacement amplitude $U$, polarisation unit vector $\mathbf e$, wavevector $\mathbf k$, angular frequency $\omega$, and initial phase $\psi_0$. Its displacement is
\begin{equation}
    \mathbf{u}(\mathbf{x},t)
    = U\mathbf{e}\,e^{i(\mathbf{k}\cdot\mathbf{x}-\omega t+\psi_0)}.
\end{equation}
For convenience, we denote the wave phase by $\Phi(\mathbf{x},t)=\mathbf{k}\cdot\mathbf{x}-\omega t+\psi_0$. We can calculate its associated strain tensor as
\begin{equation}
    \boldsymbol{\varepsilon}
    = \frac{1}{2}\bigl(\nabla\mathbf{u}
    +(\nabla\mathbf{u})^{\mathsf T}\bigr)
    = \frac{iU}{2}
    (\mathbf{k}\mathbf{e}^{\mathsf T}
    +\mathbf{e}\mathbf{k}^{\mathsf T})e^{i\Phi}.
\end{equation}
Now considering a cable with trajectory $\mathbf{r}(s)$ and unit tangent $\mathbf{t}(s)=\mathbf{r}'(s)/\lVert\mathbf{r}'(s)\rVert$,
where $s$ denotes arc length, we project the strain tensor onto the
local fibre direction to obtain the axial strain $\varepsilon_{tt}=\mathbf{t}^{\mathsf T}\boldsymbol{\varepsilon}\mathbf{t}$.
This reveals an amplitude scaling based on the projected polarisation and wave vectors as
\begin{equation}
    \varepsilon_{tt}(s,t)
    = iU [\mathbf{k}\cdot\mathbf{t}(s)]
    [\mathbf{e}\cdot\mathbf{t}(s)]
    e^{i\Phi(\mathbf{r}(s),t)}.
\end{equation}
Accounting for a gauge length $L$, the complex gauge-length-averaged axial strain measured along the fibre is
\begin{equation}\label{eq:strain_elastic_wave}
    \varepsilon_{tt}(s,t)
    =
    \frac{iU}{L}
    e^{-i\omega t+i\psi_0}
    \int_{s-L/2}^{s+L/2}
    [\mathbf{k}\cdot\mathbf{t}(\xi)]
    [\mathbf{e}\cdot\mathbf{t}(\xi)]
    e^{i\mathbf{k}\cdot\mathbf{r}(\xi)}
    \,d\xi,
\end{equation}
where $\xi$ is the integration coordinate along the gauge \citep[see also][]{egorov2018elastic,luo2020horizontally}.

We can now assume a locally rectilinear fibre around each arc-length, with constant tangent $\mathbf{t}_0$, the projected wavenumber is $k_p=\mathbf{k}\cdot\mathbf{t}_0$. In this case, gauge-length averaging introduces an amplitude factor $\operatorname{sinc}(k_pL/2)$, while incidence angle and polarisation contribute additional amplitude scaling \citep[e.g.,][]{dean2017gaugelength}. As such, we absorb these effects into an effective amplitude $A$, leading to a DAS response that retains the sinusoidal phase dependence of the incident plane wave.

In what follows,  we consider real-valued strain instead of a complex-valued phasor, since the monogenic signal is defined for real-valued functions. In addition, because the Riesz transform applied to DAS acts jointly on distance and time, its resulting orientation depends on the relative scaling of these coordinates. Here, we apply the transform directly to the sampled DAS record, using the dimensionless coordinates $x_1=s/\Delta s$ and $x_2=t/\Delta t$, where $\Delta s$ and $\Delta t$ are the channel spacing and temporal sampling interval, respectively. This scaling is analogous to the implicit scaling in conventional $f$--$k$ analysis, where the transform is computed on the sampled data directly and the sampling intervals are used to recover frequencies and wavenumbers in physical units. In the case of the Riesz transform, we explicitly write this common physical scale because the coordinates enter the norm in the frequency response. We note, however, that other choices of scaling are possible, and we discuss the implications in the Discussion.

With all these qualifications, we can write the plane wave as sensed by the locally rectilinear fibre in dimensionless coordinates as
\begin{equation}
    d(x_1, x_2)
    =
    A \cos \left(k_p \Delta s\, x_1 - \omega \Delta t\, x_2\right).
\end{equation}
For convenience define $\phi = [k_{p} \Delta s, - \omega \Delta t ] \cdot \mathbf{x}$. Hence, in the dimensionless coordinates, it becomes clear that this is the case of a two-dimensional plane wave described earlier, so that its Riesz transform is in its two components
\begin{equation}
    {\mathcal{R}}\{d\}
    =
    A \sin(\phi)
    \frac{1}{\sqrt{k_p^2 \Delta s^2 + \omega^2 \Delta t^2}}
    \begin{bmatrix}
        k_p \Delta s \\
        -\omega \Delta t
    \end{bmatrix}.
\end{equation}
Recalling the directional-Hilbert interpretation of the Riesz transform, the monogenic amplitude and phase recover exactly the plane-wave amplitude $A$ and phase $\phi$. The monogenic orientation is aligned with the phase-gradient direction in the dimensionless $(x_1,x_2)$ coordinates, or equivalently is normal to the constant-phase ridges in distance--time space. It follows that
\begin{equation}
    \tan \theta
    =
    \frac{\mathcal{R}_2 \{d \} }{\mathcal{R}_1 \{d \}}
    =
    - \frac{\omega \Delta t}{k_p \Delta s}
    =
    - \frac{1}{c_\Delta p},
\end{equation}
which shows directly that the orientation can be used to estimate the apparent slowness of the wave. The factor $c_\Delta = \Delta s/\Delta t$ is a velocity metric that converts the dimensionless orientation-derived ratio into an apparent slowness $p$ with physical units.

\begin{figure*}
    \includegraphics[width=\linewidth]{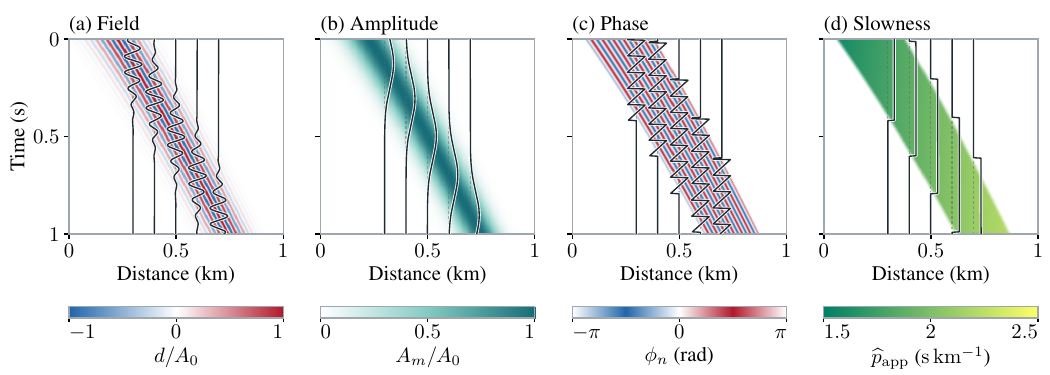}
    \caption{Monogenic analysis of a synthetic DAS wave packet. A plane wave with a Gaussian envelope normal to its wavefront is sampled along a gently curved cable, producing a localised arrival with gradually varying apparent slowness. Panels show (a) the synthetic field, (b) monogenic amplitude, (c) signed directional (monogenic) phase, and (d) estimated apparent slowness. The amplitude follows the packet envelope, the phase describes progression through the oscillatory cycle, and the slowness reflects the changing incidence angle along the cable. Overlaid traces show the temporal responses at five selected channels, with dashed lines marking their zero baselines. }
    \label{fig:example_packet}
\end{figure*}

This interpretation follows naturally from the visual inspection commonly applied to DAS records, in which the slopes of coherent ridges in distance--time space suggest the apparent propagation velocity of an arrival. The derivation formalises this visual relationship and connects the monogenic attributes directly to the seismic properties of the arrival. Figure \ref{fig:example_packet} depicts the attributes on a synthetic example of a plane wave packet arriving on a gently curved fibre, illustrating the decomposition of information.

\subsection{Robust estimation and directional coherency}

In practice, the recorded field contains various amounts of random noise, which can destabilise pointwise monogenic attributes by perturbing the estimated wavefront direction. To increase robustness, we can introduce the assumption that arrivals remain directionally coherent across neighbouring channels. As such, we estimate their orientation (and hence, the apparent slowness) over a local space–time neighbourhood defined by a non-negative weighting window $w(\mathbf{x})$ (e.g., a Gaussian kernel). 

To this end, rather than averaging the monogenic attributes directly, we adopt the structure-tensor approach proposed by \citet{unser2009multiresolution} to exploit the joint directional information contained in the Riesz vectors. For a sample point $\mathbf{x}_0$, and letting $r_1, r_2$ be the Riesz responses of the function $f$, the structure tensor is a $2\times2$ symmetric matrix $\mathbf{J}(\mathbf{x}_0) $ whose entries $m,n\in\{1,2\}$ are
\begin{equation}
    J_{mn}(\mathbf{x}_0) = \int_{\mathbb{R}^2} w(\mathbf{x} - \mathbf{x}_0) r_m(\mathbf{x})r_n(\mathbf{x}) d\mathbf{x}.
\end{equation}
 The Riesz products implicitly weight orientations by their local energy $r_1^2 + r_2^2$ while preventing cancellation when the Riesz vector reverses sign during a wave cycle. The principal eigenvector gives the dominant wavefront normal, and its eigenvalue $\lambda_{\max}$ measures the Riesz energy along that direction. Together with the smaller eigenvalue, it defines the directional coherency
\begin{equation}
    \mathcal{X} =
    \frac{\lambda_{\max} - \lambda_{\min}}
         {\lambda_{\max} + \lambda_{\min} + \tau}.
\end{equation}
The coherency approaches 1 when a single dominant orientation is present and decreases as the directional contributions become more balanced. We add $\tau\geq0$ to stabilise coherency in low-energy regions, analogous to water-level regularisation. The effect of $\tau$ is equivalent to adding isotropic energy to the tensor; concretely $\mathbf{J}(\mathbf{x}_0)_\tau = \mathbf{J}(\mathbf{x}_0) + \frac{\tau}{2}\mathbf{Id}$ where $\mathbf{Id}$ is the identity. 

\begin{figure}
    \centering
    \includegraphics[width=0.5\linewidth]{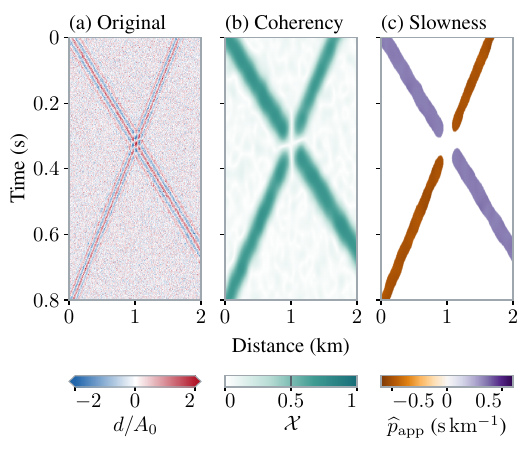}
    \caption{Robust orientation estimation for synthetic crossing wave packets. (a) Two plane-wave packets with Gaussian envelopes and apparent slownesses of $1/3$ and $-1/2\,\mathrm{s\,km^{-1}}$ are superposed and contaminated with additive Gaussian noise. (b) Directional coherency $\mathcal{X}$ computed from the locally averaged Riesz tensor, using Gaussian averaging widths of 40 m and 0.016 s. Coherency remains high along the isolated arrivals and decreases at their crossing, where competing directions contribute comparable energy, and in the noise-dominated background. (c) Apparent slowness estimated from the principal tensor orientation, retaining samples that satisfy $\mathcal{X}\geq0.50$. }
    \label{fig:crossing_coherency}
\end{figure}

Projecting the Riesz vector onto the dominant direction of the structure tensor gives the directional Hilbert response, from which instantaneous amplitude and phase follow. This approach is best suited to neighbourhoods dominated by a single wavefront orientation, and we can threshold $\mathcal{X}$ to retain directionally consistent estimates. To exemplify this, we created a synthetic crossing of wave packets with added Gaussian noise and computed the monogenic attributes through the tensor estimate, as shown in Figure~\ref{fig:crossing_coherency}. The results illustrate both the favourable settings and limitations of the method: locally coherent regions are enhanced, with robust slowness estimates, whereas crossings and regions without a dominant slowness produce unstable estimates and are removed through coherency thresholding. Equally, coherent noise forming fringes in the recordings may also be enhanced. We later show in our experiments that this thresholding enhances the information carried by coherent arrivals.

\subsection{Cable geometry bias}

For curved cable geometries, the Riesz transform is well defined, but the physical interpretation of the resulting monogenic quantities has some nuances. Because the transform kernel has infinite support decaying as an inverse power law \citep[e.g., ][]{ward2013decay}, the response at a given point contains contributions from samples over an extended portion of the fibre. If the cable is curved over this region, these samples correspond to different local projections of the arriving wavefield. In this case, the response reflects a spatially weighted combination over the effective support of the kernel, and the interpretation is no longer necessarily that of the wavefield incident at the centre point.

To assess cable geometry effects, we quantify the bias associated with a locally rectilinear approximation along the support of the kernel $\mathbf{h}(s,t)$. This  can be the Riesz kernel or its combination with a wavelet pre-filter. Intuitively, because the kernel is decaying, each point futher away contributes less to the response. As such, we can expect the bias to depend on the departure of the cable from its centre tangent. 

As derived in section~\ref{sec:monogenic_das}, the apparent slowness at point $s_0$ in the rectilinear cable with projected wavenumber $k_p = \mathbf k\cdot\mathbf t(s_0)$ is just $p(s_0) = \frac{k_p}{\omega}$. We consider then the apparent slowness estimate as projected onto the real three-dimensional cable $\widehat p_{\mathrm c}(s_0)$, which is computed from its Riesz vector applied to the synthetic DAS field generated on the true cable using the axial-strain forward model from Equation \ref{eq:strain_elastic_wave}.  Hence, we can define the apparent slowness bias with respect to $\widehat p = p(s_0)$ at every point along the fibre to be 
\begin{equation}
    B_p(s_0;\mathbf k,\omega)
    =
    \frac{
        \left|
            \widehat p_{\mathrm c}
            -
            \widehat p
        \right|
    }{
        \left|
            \widehat p
        \right|
    } .
\label{eq:geometry_slowness_bias}
\end{equation}
The bias in the axial-strain model is also dependent on parameters such as polarisation and gauge length which we assume to be fixed. Small values of $B$ indicate that the Riesz response of the curved-cable record remains close to that of the ideal locally rectilinear cable having the same phase and apparent wavenumber at the centre point.

\begin{figure}
    \centering
    \includegraphics[width=0.5\linewidth]{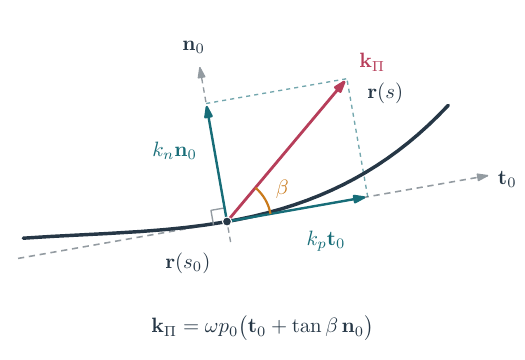}
    \caption{Wave vector decomposition for a planar curved cable. At $\mathbf r(s_0)$, the local tangent $\mathbf t_0$ and in-plane normal $\mathbf n_0$ resolve the projected wavevector as $\mathbf k_\Pi=k_p\mathbf t_0+k_n\mathbf n_0$. The incidence angle $\beta$ gives $k_n=k_p\tan\beta$, with $k_p=\omega p_0$. }
    \label{fig:cable_geometry}
\end{figure}

This bias can be evaluated numerically for a known cable trajectory $\mathbf r(s)$ for any test plane wave with $(\mathbf k,\omega)$. To parameterise the test wavevectors relative to the cable geometry, we consider a planar cable lying in the plane $\Pi$, with local tangent $\mathbf t_0$ and in-plane normal $\mathbf n_0$ at $s_0$. Only the projection $\mathbf k_{\Pi}$ of the three-dimensional wavevector onto this plane contributes to the phase variation sampled along the cable.  Defining the local apparent slowness as $p_0=(\mathbf k\cdot\mathbf t_0)/\omega$ gives the tangential wavenumber $k_p=\omega p_0$. If $\beta$ denotes the in-plane angle between the projection $\mathbf k_\Pi$ and $\mathbf t_0$, then the in-plane normal component is $k_n=k_p\tan\beta$, such that the relevant projected wavevector may be written $\mathbf k_\Pi=\omega p_0(\mathbf t_0+\tan\beta \, \mathbf n_0)$. This geometry is depicted in Figure~\ref{fig:cable_geometry}. 

\begin{figure*}
    \includegraphics[width=\linewidth]{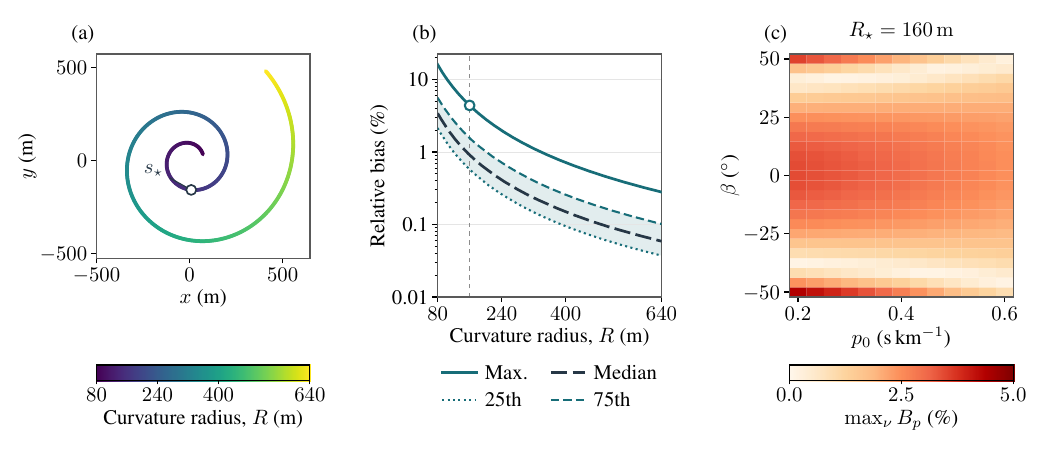}
    
    \caption{Geometry-induced apparent-slowness bias on a planar logarithmic spiral. (a) Analysed cable segment coloured by its radius of curvature $R$, which increases monotonically with arc length. (b)  Relative bias $B_p$ against the locally straight-cable reference at each channel, shown on a logarithmic ordinate. The sweep comprises planar P-waves at 8, 12 and 16 Hz, with $p_0=0.2$--$0.6\,\mathrm{s\,km^{-1}}$ and $\beta=-50^\circ$--$50^\circ$. The lines indicate the maximum and different percentiles of bias in the sweep. (c) Bias as a function of $p_0$ and $\beta$ at the marked point, $R_\star=160\,\mathrm m$, maximised over frequency. Slowness is estimated from the cycle-averaged Riesz tensor of gauge-averaged axial strain, using a 10 m gauge length, 5 m channel spacing and 5 ms time sampling. }
    \label{fig:spiral_geometry_bias}
\end{figure*}

With the change of variables, the geometry-induced bias can be parameterised in a frequency--slowness--incidence domain as $B_p(s_0;\omega,p_0,\beta)$, where $\omega$ controls the temporal scale, $p_0$ is the reference apparent slowness at the centre point, and $\beta$ controls the transverse incidence responsible for sensitivity to cable curvature. We perform a synthetic experiment to characterise the curvature bias, shown in Figure~\ref{fig:spiral_geometry_bias}. We choose a curved cable following a logarithmic spiral geometry, in which the radius of curvature $R$ increases linearly with the cable arc-length $s_0$. For every point along the cable, we sweep a range of planar-wave parametres $(\omega,p_0,\beta)$, with fixed gauge length and P-wave polarisation, and record the maximum bias within the sweep. In ~\ref{fig:spiral_geometry_bias}b, it is shown that as the cable becomes more locally rectilinear, the bias decreases from about 15\% at 80 metres radius to 0.3\% at 600 m radius. In panel~\ref{fig:spiral_geometry_bias}c, we decompose the bias into the incidence and tangent apparent slowness coordinates, showing how the bias decreases for larger $p_0$. The bias is shown to decrease at $\sim40^\circ$ and increase again. This occurs because the signed slowness error crosses zero, that is, the estimate changes from underestimating to overestimating slowness. We refer the reader to \citet{WangFichtner2026GeometricFiltering} for further details on the effects of planar fibre curvature.

\subsection{Adaptive structure-tensor window}

The structure-tensor estimates and associated directional coherency introduced above are important parts of monogenic analysis in real recordings. Nevertheless, because these estimates require computation in a neighborhood of samples, their weighting window should remain within a region where the projected slowness is sufficiently consistent. 

Using the planar model above (Figure~\ref{fig:cable_geometry}), because $p(s)=(\mathbf k\cdot\mathbf t(s))/\omega$, we have that $p(s_0+u)/p_0=\cos\Delta\vartheta+\tan\beta\sin\Delta\vartheta$, where $\Delta\vartheta=\vartheta(s_0+u)-\vartheta(s_0)$ is the change in tangent angle. Consequently, the cable turning and the incident angle $\beta$ are the two factors informing the appropriate window width. For example, when $\beta$ approaches 90$^\circ$, the incidence becomes perpendicular and hence weakly sensed. In that case, small bends will affect more strongly than if incidence was closer to the cable tangent. 

However, many times, the incidence angle is the unknown, so we need to select a width without knowing the wave propagation direction.  To do so, we introduce a geometry-based heuristic to adapt the width of the Gaussian window to a rectilinear support, based on the cumulative turning angle as we move along the cable from the analysed centre point. Specifically, writing $\kappa(s)=d\vartheta/ds$, we accumulate absolute turning as $\int_{s_0}^{s_0+u}|\kappa(v)|\,dv$ for $u>0$, and
similarly in the opposite direction. We extend the interval equally on both
sides until either accumulated turn reaches a maximum angular tolerance $\Theta_{\max}$. Then, we take the resulting half-length as $3\sigma_s$, additionally subject to a maximum width $\sigma_{\max}$.

\begin{figure*}
    \includegraphics[width=\linewidth]{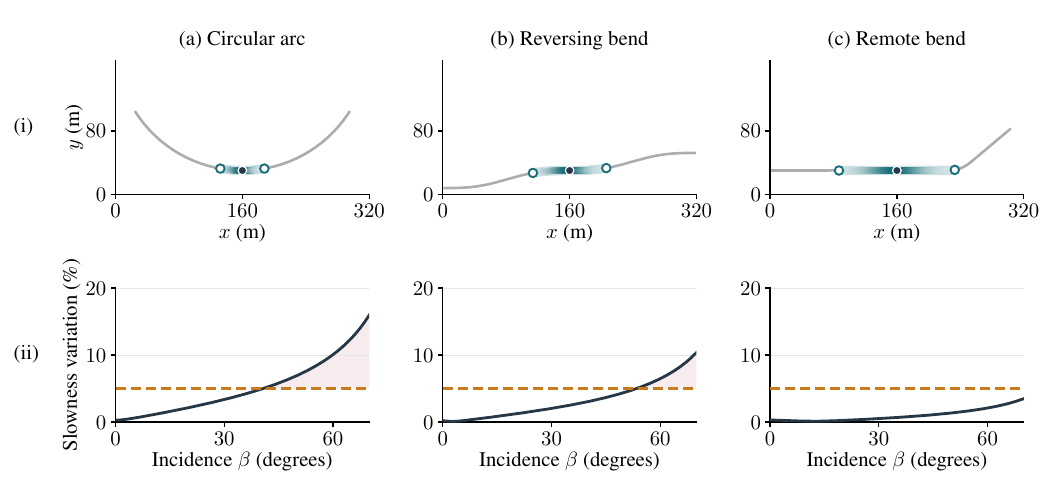}
    \caption{Geometry-adaptive spatial averaging for (a) a circular arc, (b) a reversing bend and (c) a remote bend. (i) Cable geometry with the centre marked by a dark dot and the geometry-selected neighbourhood in teal. Open circles lie three Gaussian standard deviations from the centre in each arc-length direction. (ii) Gaussian-weighted relative slowness variation at the geometry-selected width as a function of incidence; the dashed amber line marks the 5\% tolerance.}
    \label{fig:geom_adaptive_heuristic}
\end{figure*}

This heuristic is showcased in Figure~\ref{fig:geom_adaptive_heuristic}(i)
for three representative cable geometries, using $\Theta_{\max}=10^\circ$
and $\sigma_{\max}=80$~m. Figure~\ref{fig:geom_adaptive_heuristic}(ii) shows the Gaussian-weighted RMS relative slowness variation about the centre value for these heuristic widths, with varying incidence angle. The dashed line marks a 5\% RMS tolerance for comparison. As incidence approaches perpendicular to the cable, the projected component becomes weaker and small bends can cause larger relative variations. In those cases, a smaller turning angular tolerance may be needed.

\section{Data}

\begin{figure}
    \centering
    \includegraphics[width=0.5\linewidth]{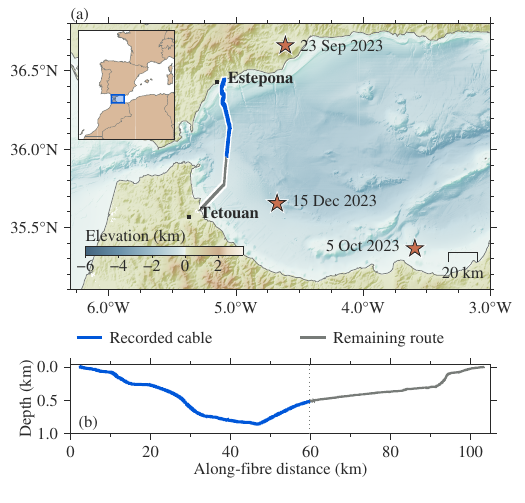}
    \caption{(a) Location of the Estepona--Tetouan fibre-optic cable in the western Alboran Sea. The DAS-interrogated section is shown in blue and the remaining cable route in grey. Stars mark the epicentres of the three earthquakes considered in this study, with origin times in UTC: 23 September 2023 at 04:22:00 ($m_b=3.2$), 5 October 2023 at 09:00:45 ($m_{bLg}=3.5$), and 15 December 2023 at 23:45:11 ($m_{bLg}=3.4$). The inset locates the study area. (b) Water depth along the cable, with the dotted line marking the end of the interrogated section.}
    \label{fig:data_maps}
\end{figure}

To illustrate the applications of our monogenic analysis methodology, we use a submarine DAS deployment in the western Alboran Sea. The dataset was acquired on the Telxius Estepona-Tetouan telecommunication cable \citep{ugalde2023safe}. An approximately 60-km-long section of the cable, reaching a maximum water depth of 860 m, was interrogated between 21 September 2023 and 10 February 2024 (Figure~\ref{fig:data_maps}). A chirped-pulse HDAS interrogator manufactured by Aragón Photonics was used in the deployment, with a gauge length and spatial sampling interval of 10 m. It includes data sampled at 100 and 250 Hz, depending on the acquisition period. The three earthquakes considered in the applications are shown in Figure 7a. Their origin times, epicentral locations and magnitudes were obtained from the earthquake catalogue of the Instituto Geográfico Nacional \citep{IGN2022}. The earthquake of 5 October 2023 is analysed in Section 4, while the other two events are presented in the Supplementary Material.

\section{Applications}

 We consider the \textit{T}-wave trains recorded following the $m_{bLg} \ 3.5$ earthquake of 5 October 2023. These arrivals exemplify how the different attributes can be used in conjunction. Using directional coherency, travelling patterns are enhanced. Monogenic amplitude and phase are used to identify individual oscillations, and monogenic orientation provides the apparent-slowness estimates at each point, from which an incoming propagation direction can be inferred.

\subsection{Directionally coherent signal enhancement}

\begin{figure*}
    \includegraphics[width=\linewidth]{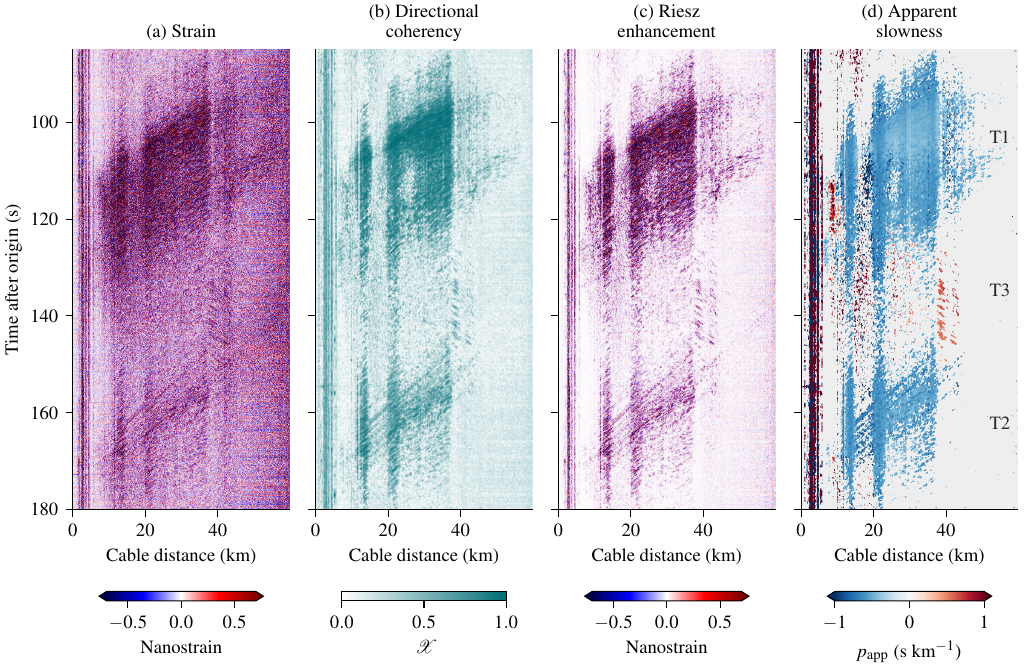}
    \caption{Directional enhancement of the 5 October 2023 \textit{T}-wave trains. (a) Strain filtered at 3--30 Hz. (b) Directional coherency. (c) Reconstruction after continuous coherency weighting of the monogenic components, shown on the same amplitude scale as (a). (d) Riesz apparent slowness where $\mathcal{X}\geq0.3$, computed from the input strain. Time is measured from earthquake origin. We identify T1, T2 and T3 for analysis later.}
    \label{fig:denoising_slowness}
\end{figure*}

\textit{T}-waves provide a favourable example because their extended oscillatory trains form coherent ridges in distance--time space. As a first important demonstration of the monogenic analysis, we show that the directional coherency of the Riesz tensor can enhance the parts of the field that locally, contain only a single predominant orientation.  

The strain data were band-pass filtered between 3 and 30~Hz and the the spatial median across channels was subtracted from each channel at each time sample. This recording has a temporal sampling interval of 0.01~s and a channel spacing of 10~m, which gives a coordinate scale for the Riesz transform of $c_\Delta=1000\ \mathrm{m\,s^{-1}}$. The Riesz components are computed using a single FFT of the full distance-time matrix using reflection padding. We obtain robust estimates and coherency through a structure-tensor using a separable Gaussian window with standard deviations of 90~m and 0.09~s respectively, which stay well within our curvature-bias criterion, and we set $\tau=0.01\ \mathrm{nanostrain}^2$ for regularising the coherency estimates. 

With these ingredients, we enhance directionally coherent patterns by multiplying the three monogenic components by the directional coherency $\mathcal X(s,t)$, which provides a spatio--temporal weight that emphasizes the coherent parts of the monogenic attributes. As the monogenic representation is higher-dimensional, its components are not independent, hence their weighting need not correspond exactly to the monogenic representation of a single strain field. Thus, we obtain the enhanced strain
in the least-squares sense, corresponding to the strain
field whose monogenic components best match the weighted ones. We provide the derivation of the enhanced strain in Appendix~\ref{app:monogenic_enhancement}

Figure~\ref{fig:denoising_slowness} exhibits how the two main \textit{T}-wave trains, denoted T1 and T2, emerge from the background after coherency-weighting. Their oscillations remain visible on the same strain scale, while much of the less directionally coherent background is attenuated. The apparent-slowness map follows the moveout of these trains over extended cable sections, retaining spatial variations that are much less evident in the original strain waveform display, revealing the existence of arrivals with contrasting propagation (denoted T3). The displayed slowness is computed from the input strain, with a $\mathcal X\geq0.3$ threshold. Other coherent features remain visible near the land–sea transition at 2.39 km along the cable, possibly associated with breaking waves and exhibiting apparent slownesses distinct from those of the main \textit{T}-wave trains. An additional wave train with apparent slowness of opposite sign is also visible during intervals when the main \textit{T}-wave trains are absent. We return to this arrival later in the text, using direction-selective enhancement to examine it more closely.

\subsection{Cycle detection and phase association}

\begin{figure*}
    \includegraphics[width=\linewidth]{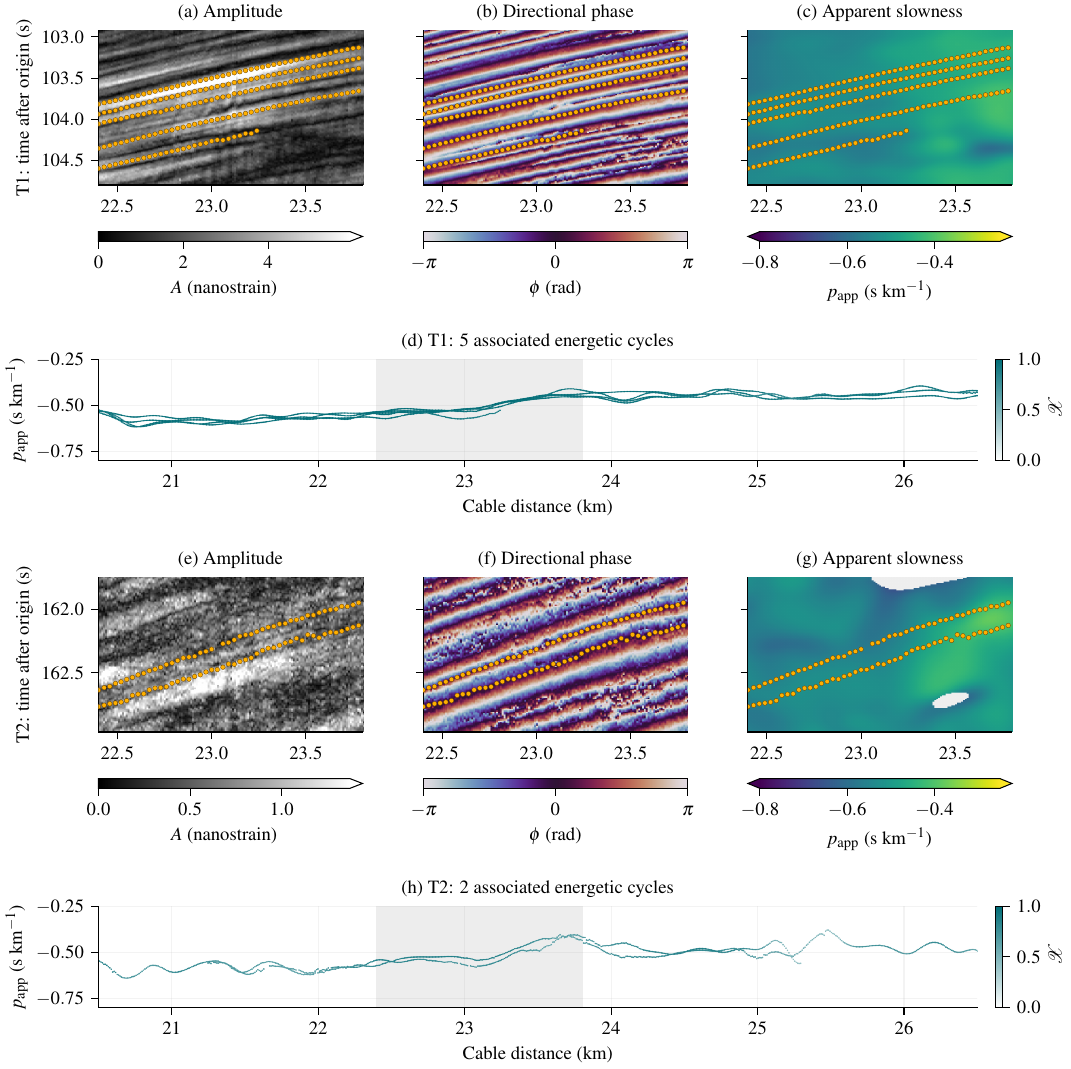}
      \caption{Phase-assisted association within the \textit{T}-wave packets T1 (a--d) and T2 (e--h) recorded along the Estepona--Tetouan cable. Maps show monogenic amplitude, directional phase and apparent slowness. Orange dots mark associated equal-phase crossings. Panels (d,h) show local median slowness for each associated oscillation and channel, coloured by directional coherency. Grey bands locate the map close-ups.} 
  \label{fig:phase_association}
\end{figure*}

Next, we use the monogenic amplitude, phase and orientation together to follow individual oscillations within the two main identified \textit{T}-wave trains, T1 and T2, which are selected within the time windows of 90--150~s and 150--200~s after earthquake origin, respectively.

First, we use the monogenic amplitude and directional coherency to
locate an energetic, coherent portion of each train.  Its maximum within each train's search window provides a reference time of 103.55 s for T1 and 162.60 s for T2. 

Then, we search within $\pm6$ s of each reference time for individual
oscillations. Candidate cycles are identified from positive-going
zero crossings of the signed directional (monogenic) phase, computed along the principal Riesz-vector direction. Each selected phase crossing is then followed in both cable directions. If $t_i$ is its time at channel $i$, the local orientation predicts its continuation to a neighbouring channel through
\begin{equation}
    t_{i+1}^{\mathrm{pred}}
    =
    t_i
    +
    \frac{p_{\mathrm{app},i}+p_{\mathrm{app},i+1}}{2}
    (s_{i+1}-s_i),
    \label{eq:phase_transport}
\end{equation}
with slowness given by the monogenic orientation. At each step, we first exclude candidate crossings with directional coherency $\mathcal{X}<0.4$. Among the remaining candidates, we choose the one closest to the prediction, requiring a correction smaller than 0.35 of the local period, with tolerance of half a temporal sample. Crossing times are interpolated between temporal samples, and the local period is estimated from neighbouring crossings. A track stops if it cannot bridge a gap of at most 30 m.

Figure~\ref{fig:phase_association} illustrates this process, showcasing the complementary information carried by the monogenic attributes. Amplitude describes where the oscillations are strong, while phase resolves their progression through a cycle, and the orientation-derived slowness describes their changing moveout. The nearby oscillations of T1 give similar spatial slowness profiles, while T2 shows a weaker signal with greater variability and interrupted associations.

\subsection{Apparent slowness to direction of propagation} 

We demonstrate the reliability of the proposed analysis through a downstream task. Using the known cable locations, the associated apparent slownesses, as computed in the previous section, can be used to estimate an incoming direction of the \textit{T}-wave trains.  We use the 21--26 km section of the Estepona--Tetouan cable, whose change in tangent direction provides complementary projections of the incident wavefield. The fit is performed separately for T1 and T2.

\begin{figure*}
    \includegraphics[width=\linewidth]{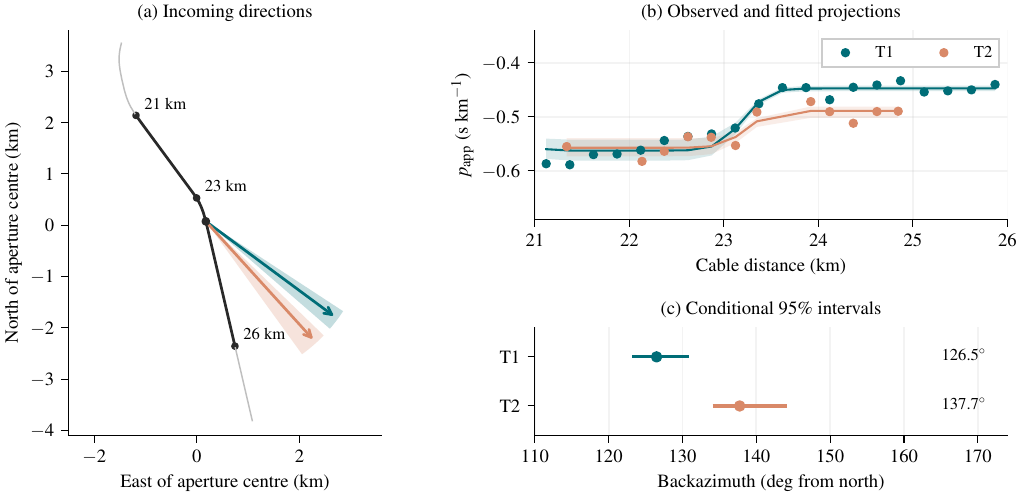}
    \caption{Incoming directions inferred from phase-associated Riesz slowness over the 21--26 km cable section offshore Estepona. (a) Estimated backazimuths and angular intervals, anchored at the aperture midpoint. (b) Observed spatial-bin medians and fitted cable projections. (c) Backazimuth estimates and 95\% bootstrap intervals. Shading represents bootstrap intervals conditional on the selected oscillations.}
    \label{fig:direction_fits}
\end{figure*}

For this calculation, samples within the associated phase neighbourhoods  are retained only where $\mathcal X\geq0.7$. We first take their median within each cycle, and then a channel median across cycles. To avoid treating densely spaced, correlated channels as independent observations, we divide the aperture into 250 m bins. Each bin requires at least five accepted channels and contributes its median slowness and the mean cable tangent of those channels. The surveyed geometry is smoothed to avoid artificial sharp transitions when calculating the three-dimensional unit tangent.

The predicted along-fibre apparent slowness in bin $b$ can be approximated as
\begin{equation}
    p_{\mathrm{app},b}^{\mathrm{pred}}
    =
    \overline t_{E,b}\,p_E
    +
    \overline t_{N,b}\,p_N ,
    \label{eq:horizontal_projection}
\end{equation}
where $(p_E,p_N)$ are the east and north components of the horizontal slowness vector, and $(\overline t_{E,b}, \overline t_{N,b})$ the respective components of the averaged tangent in the bin. The implications of neglecting the vertical contribution $\overline t_{U,b}\,p_U$ are discussed in section 5.2. 

We estimate $(p_E,p_N)$ by minimising the sum of Huber losses of the bin residuals, with equal weight per bin. The Huber loss \citep{Huber1964} is defined as
\begin{equation}
    \rho_\delta(r)=
    \begin{cases}
        \frac{1}{2}r^2, & |r|\leq\delta,\\
        \delta\left(|r|-\frac{\delta}{2}\right), & |r|>\delta,
    \end{cases}
\end{equation}
where $\delta$ is the transition scale, here set to
$0.03\ \mathrm{s\,km^{-1}}$. Thus, the loss is quadratic for small residuals and linear
for larger ones, reducing the influence of individual discrepant bins.

The results are shown in Figure~\ref{fig:direction_fits}, where the inferred vector points into the incoming backazimuth. To assess sensitivity to spatial sampling, we resample pairs of slowness and tangent in contiguous 500 m blocks, without bridging gaps between accepted bins, and repeat the fit for 400 bootstrap draws. The central 95\% angular range of the accepted fits is shown in the figure. These intervals are conditional on fixed parametres and assumptions that include the selected oscillations, coherency criterion, geometry and horizontal-wave approximation.

\begin{figure}
    \centering
    \includegraphics[width=0.5\linewidth]{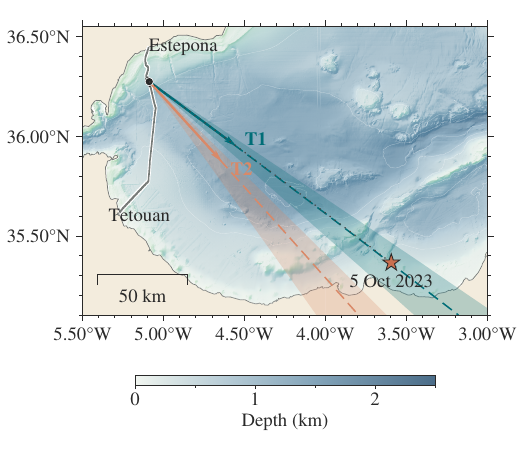}
    \caption{Estimated incoming directions of T1 and T2 over western Alboran Sea bathymetry. Coloured cones show the 95\% bootstrap angular intervals projected from the midpoint of the fitted cable section. The star marks the earthquake epicentre, with a dotted reference bearing from the cable.}
    \label{fig:direction_cones}
\end{figure}

The fitted projections reproduce the main spatial change in apparent slowness as the cable turns near 23 km. T1 gives an incoming backazimuth of of $126.5^\circ \pm 3.9^\circ$, and T2 gives $137.7^\circ \pm 5.0^\circ$, with uncertainties calculated via bootstrap.

Figure~\ref{fig:direction_cones} projects these incoming directions over the regional bathymetry from the cable point at 23.5 km. The T1 sector includes the bearing towards the catalogue epicentre, whereas T2 has a more southerly incoming direction, which is consistent with the two selected packets sampling different propagation paths. These results demonstrate that the local Riesz estimates can supply spatially consistent projections for a subsequent directional analysis.

\subsection{Direction-selective enhancement of a weaker arrival}

\begin{figure*}
    \includegraphics[width=\linewidth]{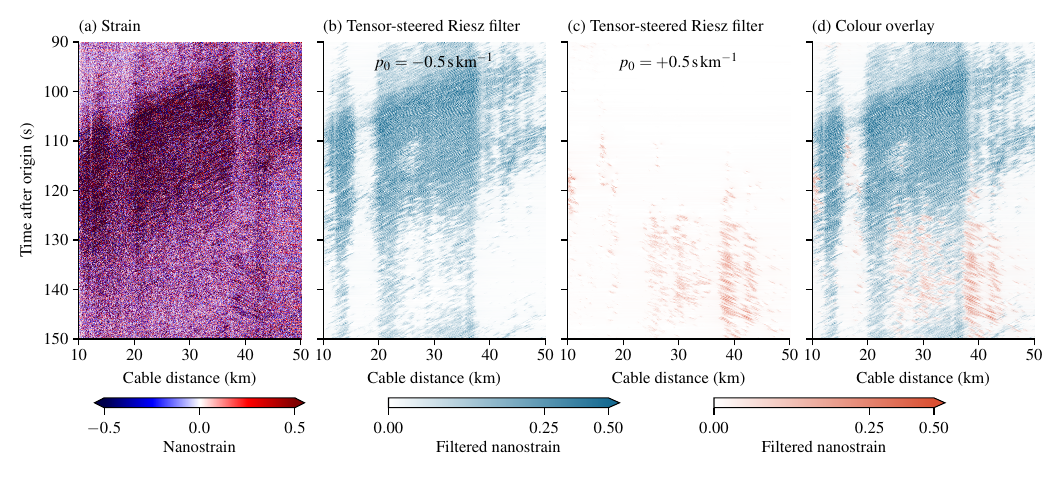}
    \caption{Direction-selective Riesz enhancement of the 5 October 2023 \textit{T}-wave  recording in the western Alboran Sea. (a) Preprocessed strain. (b,c) Tensor-steered reconstructions for reference slownesses of $-0.5$ and $+0.5\,\mathrm{s\,km^{-1}}$, respectively. (d) Colour overlay of both selections. Panels (b--d) display positive strain lobes only for clearer visualisation.}
\label{fig:direction_selection}
\end{figure*}

Finally, we extend the directional enhancement to distinguish contributions with contrasting moveout. The important observation is that the Gaussian window used to average the Riesz products can be elongated and rotated to follow a reference moveout, $\Delta t=p_0\Delta s$, averaging over a longer neighbourhood along the wavefront and a shorter one across it. This idea follows the broader use of anisotropic neighbourhoods in structure-tensor analysis \citep{brox2006adaptivestructure}. We use Gaussian standard deviations of 300 and 50 m along these two directions in the scaled space--time coordinates, and consider reference slownesses $p_0=\pm0.5\,\mathrm{s\,km^{-1}}$. The window remains centred on each sample, and the Riesz transform itself is unchanged. Note that rotating the averaging window does not reject other propagation directions, but can be useful to avoid mixing different contributions in the tensor window.

Then, we can introduce an explicit selection by combining directional coherency with agreement with the reference orientation. Using the same reconstruction framework as earlier, we apply the continuous weight $g=\mathcal{X}\exp[-\Delta\theta^2/(2\sigma_\theta^2)]$, where $\Delta\theta$ is the smallest angle between the estimated wavefront normal and the reference normal, treating orientations separated by $180^\circ$ as equivalent. We set $\sigma_\theta=10^\circ$ and retain a $\tau=0.01\,\mathrm{nanostrain}^2$ in the coherency calculation. Figure~\ref{fig:direction_selection} shows how the two selections emphasise contrasting moveouts, clearly revealing weaker opposite-slope patterns beneath the dominant train. 

\begin{figure}
    \centering
    \includegraphics[width=0.5\linewidth]{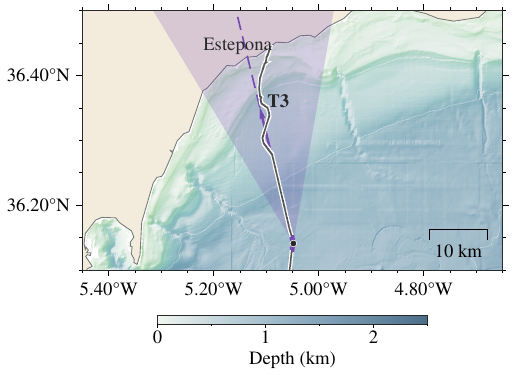}
    \caption{Incoming direction of the weaker, opposite-going T3 arrival over bathymetry near Estepona. The arrow shows the fitted backazimuth and shading its conditional 95\% spatial-block bootstrap interval, anchored at the fitted cable aperture. The estimate uses phase-associated data with 300m Ricker-wavelet spatial filtering. The sector indicates a possible northern-margin reflection contribution.}
    \label{fig:t3_cone}
\end{figure}

We denote the opposite-slope train as T3 and analyse its propagation direction. We associate five oscillations near 135 s at the 39 km reference channel, retaining slowness samples with $\mathcal X\geq0.4$, with phase neighbourhoods containing opposite-sign estimates excluded. The estimates give an incoming backazimuth of approximately $346^\circ$, with a conditional 95\% bootstrap interval of $329^\circ$--$10^\circ$, crossing north. Figure~\ref{fig:t3_cone} shows that this sector points towards the northern Alboran margin near Estepona, compatible with a reflected or scattered contribution. With this example, we illustrated how the monogenic attributes can help isolate and characterise weaker contributions, obfuscated in overlapping wavefields and with little background contrast.

\section{Discussion}

\subsection{On parameter choices and interpretation}

So far, we showcased the monogenic analysis methodology through an example containing \textit{T}-wave trains. Next, we discuss the relevant parameter choices, from the computation of the Riesz transform to the estimation of orientation, apparent slowness and directional coherency.  

\subsubsection{Fourier transform computation}

We computed the Riesz transform using a full-record two-dimensional FFT, followed by multiplication by the two component-wise frequency responses, and inverse transformation to obtain the Riesz vector field. This avoids the repeated boundaries and tapering introduced by short-window analyses, which can produce spectral leakage and reduce frequency and wavenumber resolution. We note that full-record computation does not assume that one orientation describes the whole record, because the direction is still estimated locally along the support of the Riesz kernel. The computational cost of the full method is dominated by the FFTs and local tensor averaging, with little overhead beyond these standard operations. Thus, our proposed analysis is suitable for real-time applications. 

\subsubsection{Velocity metric}
\label{sec:vel_metric}
Using the native sampling in the FFT defines the space--time metric $c_\Delta=\Delta s/\Delta t$, here 1000~m~s$^{-1}$. Because distance and time represent different physical quantities, we could instead choose another of velocity metric $c_*=\alpha c_\Delta$, leading to a change of the relative scale of the two axes. Mathematically, for an ideal plane wave, the scaled coordinates will stretch the orientation calculation, but the resulting recovered amplitude, phase and correctly converted slowness will remain unchanged. Nonetheless, when several arriving contributions overlap, a change of the metric risks favouring a particular tensor direction and increasing coherency artificially, without better resolving the wavefield. In our tests, we find that varying $c_*$ from 500 to 2000~m~s$^{-1}$, with fixed tensor widths and extraction windows, changed the median T1 apparent slowness only from $-0.469$ to $-0.476$ s/km. The less coherent T2 packet showed a larger change, from $-0.475$ to $-0.555$ s/km. As such, high coherency that persists alongside stable estimates across the sweep supports the interpretation of a consistent arrival. This reinforces our recommendation of using the sampling metric as a first choice.  

\subsubsection{Structure tensor size}
Another important choice is the spatial and temporal extent of the structure tensor. Larger windows stabilise the estimates, but can combine different cable projections across bends or mix neighbouring arrivals. Instead, smaller windows preserve these changes at the expense of greater variability, where a noisy Riesz vector can produce high coherency by chance. We used Gaussian standard deviations of 90~m and 0.09~s for the Alboran example. We also tested that independently varying the spatial standard deviation from 30 to 180~m changed the median T1 slowness by about 0.018~s~km$^{-1}$, while varying the temporal standard deviation from 0.03 to 0.2 s changed it by about 0.005~s~km$^{-1}$. As such, our proposed geometry adaptation addresses the spatial trade-off, while the temporal width should be chosen to reflect how long one orientation remains representative. We note that the tensor window can span several oscillations because it is only the directional estimate that is affected through the averaging process. The monogenic amplitude is unchanged and the directional phase is likewise evaluated at each sample from the local components, using the estimated direction (that is, the slowness).

\subsubsection{Regularisation}
Finally, the regularisation parameter $\tau$ reduces coherency where the tensor energy is small, helping reduce random positives in the background. Its units are those of squared signal amplitude, so its value should be considered relative to the energy of the background and the weakest arrivals of interest after preprocessing. We used $\tau=0.01$ nanostrain$^2$ for the main estimates. Larger values can mask weaker arrivals. For example, increasing it to 0.3 nanostrain$^2$, at a coherency threshold of 0.6, retained approximately 99.5\% of the selected T1 samples but none of the selected T2 samples. Hence, a value useful for highlighting a strong packet can also suppress a weaker arrival. Also important to highlight is that $\tau$ does not change the unmasked orientation or slowness, only their coherency and subsequent weighting.

Further numerical details and sensitivity figures are provided in the supplementary material.

\subsection{On \textit{T}-wave enhancement and analysis}

The dense spatial sampling of DAS has motivated the development of methods to enhance coherent seismic signals while suppressing background noise. These include coherency-based approaches and moving-rank-reduction filtering \citep{Chen2023Enhancing}, as well as deep-learning methods ranging from supervised convolutional networks \citep{Yang2023Denoising} to self-supervised approaches that do not require noise-free training data \citep{VanDenEnde2023SelfSupervised, Lapins2024DASN2N}. More recently, \citet{Shi2025Denoising} applied self-supervised masked autoencoders to offshore DAS recordings, improving seismic phase picking and earthquake detectability. These developments are especially relevant to submarine DAS observations, where instrumental and environmental noise can obscure T-wave arrivals. For example, \citet{atterholt2022wavefieldpartitioning} used curvelet-based wavefield partitioning to suppress stochastic and coherent noise, and \citet{shen2024oceanbottom} later developed a curvelet-denoising scheme for submarine DAS recordings that enhanced \textit{T}-wave signals and improved the detection of oceanic earthquakes.

Our approach goes beyond conventional denoising by using the local characteristics of the recorded wavefield to guide its enhancement. Specifically, we use directional coherency to selectively emphasise contributions with a consistent orientation, without relying on the suppression of particular frequency bands. This selection is performed locally in the spatio--temporal field, with the monogenic representation providing an additional characterisation through amplitude, phase and apparent slowness, which have direct physical interpretations. These attributes can then be used to associate individual oscillations and investigate the propagation of the recorded wavefield. 

This is particularly relevant for weak arrivals, where amplitude alone may not be sufficient to distinguish a signal from the background. The differences between T1 and T2 illustrate this limitation, where the weaker T2 train provides fewer continuous phase associations, making its apparent-slowness estimates more sensitive to the selected oscillations. In addition, our proposed direction-selective enhancement offers a way to examine signals that are difficult to recognise in the original recordings. For example, the T3 arrival becomes more visible after selecting the corresponding orientation.

Along a curved fibre, the apparent slowness of a wave varies with the local cable direction. \citet{atterholt2022wavefieldpartitioning} noted that this variation can complicate filtering based on apparent velocity. On that matter, our approach estimates slowness locally, so the cable needs to be approximately straight only over the effective spatial support of the Riesz kernel and structure-tensor. In fact, over a longer section, changes in cable orientation are desired, as they provide complementary projections of the incoming wavefield that can be used to estimate its backazimuth.

Applying backazimuth estimation to the Alboran recording, we find that the inferred direction of T1 agrees with the earthquake epicentre, whereas T2 has a more southerly backazimuth. This difference may reflect seismic-to-acoustic conversion at different locations or different acoustic propagation paths. Remarkably, the later T3 arrival has an incoming direction towards the Estepona coast, consistent with a possible coastal reflection of T1. A more detailed analysis of the conversion locations and the possible coastal reflection, including travel-time constraints, is currently in progress.

In our backazimuth estimation, the horizontal projection neglects the vertical slowness contribution \(p_U\sin\gamma\), where \(\gamma\) is the cable inclination. In the analysed section, the median inclination is approximately 1.3°, reaching a maximum of 2.3°. Assuming a locally plane acoustic wave propagating at 1.5 km/s, the estimated horizontal slownesses of approximately 0.57–0.58 s/km imply vertical slownesses of approximately 0.32–0.35 s/km. We evaluate the sensitivity of the inferred directions to this assumed vertical neglected contribution, and find that it reaches at most 0.012 s/km, resulting in backazimuth changes below 0.7°. These results indicate limited systematic error associated to the neglected term.

Our methodology can also be applied to other seismic and nonseismic signals recorded by DAS. In all cases, the interperation of the results is bound to the characteristics of the recording. For example, direct \textit{P}-waves arriving at steep angles to a nearly horizontal cable may produce relatively weak strain signals, while scattered waves can become more prominent in the recordings. \citet{vandenEnde2021beamforming} showed that the directional sensitivity of DAS and the sensitivity of strain measurements to local heterogeneities can strongly affect waveform coherence and complicate conventional array processing. In such cases, the dominant monogenic estimates may represent scattered arrivals instead of the direct wave.

\subsection{On multiscale analysis}

The combination of the Riesz transform with wavelets has been extensively studied, because they provide a multiscale monogenic representation allowing amplitude, phase and orientation to be examined at different scales \citep[][]{unser2009multiresolution}. The choice of wavelet also affects the localisation of the resulting Riesz components through its decay and vanishing moments \citep{ward2013decay}. For DAS, this offers a natural way to examine contributions that may be difficult to distinguish in the broadband field, where spatial and temporal filtering can be applied separately.

In our tests, spatial filtering revealed weaker \textit{T}-wave patterns that were less visible in the broadband field. However, it also changes the spatial-frequency content contributing to the tensor estimate. For an isolated plane wave retained by the filter, the apparent slowness remains unchanged. When several contributions overlap, filtering changes their relative energies and may therefore favour a different tensor direction. The resulting estimate characterises the filtered wavefield and need not coincide with that obtained from the broadband analysis.

\begin{figure}
    \centering
    \includegraphics[width=0.5\linewidth]{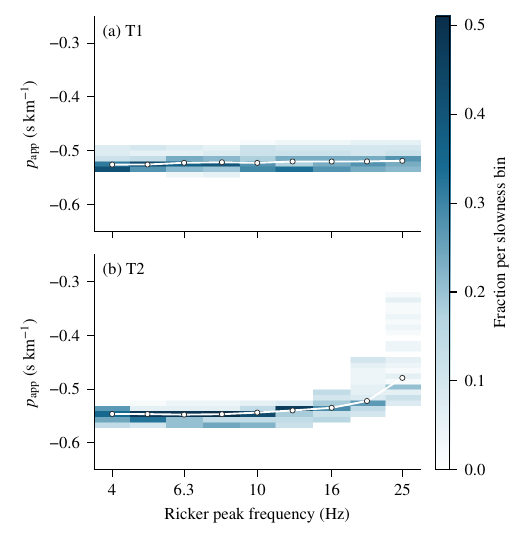}
    \caption{Apparent-slowness distributions across temporal Ricker scales for T1 (a) and T2 (b), over the 22.75--23.25 km section of the cable. Colours show normalised histograms of channel-median estimates, with white dots indicating their medians. The phase-associated samples and tensor settings are held fixed across scales.}
    \label{fig:frequency_slowness}
\end{figure}

Similarly, temporal frequency bands can be analysed. We can apply a temporal Ricker wavelet and compute the monogenic attributes and tensor estimates for the resulting field. Repeating this calculation over several wavelet scales allows us to represent the distribution of apparent slowness against temporal frequency for a selected wave packet and cable section.  As such, agreement across bands supports a consistent arrival, while differences identify frequency-dependent structure that requires a detailed interpretation. We calculate the apparent slowness across temporal Ricker scales in Figure~\ref{fig:frequency_slowness}, where we observe agreement across frequency scales in T1 and most of T2, where the disagreement appears at higher frequencies where the arrival energy becomes weaker.

\subsection{On monogenic description of perturbations}

\begin{figure*}
    \includegraphics[width=\linewidth]{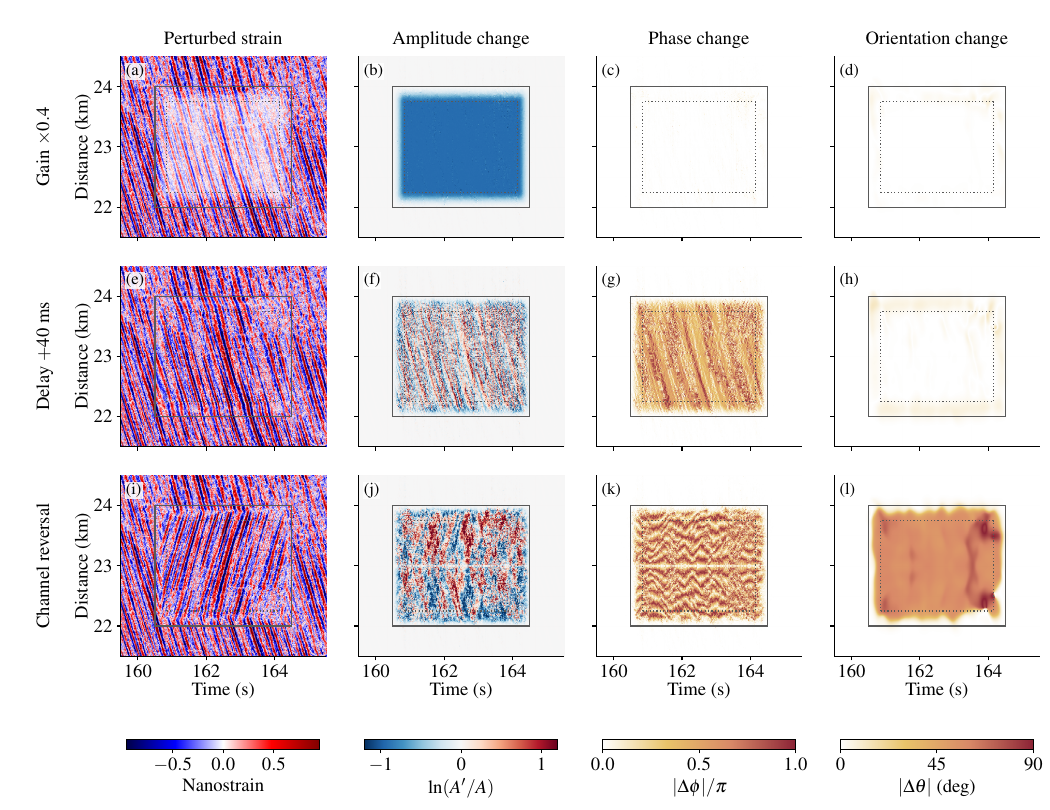}
    
      \caption{Monogenic signatures of localised perturbations to the T2 train near 23 km. Three different perturbations are studied. In each row, in order: amplitude attenuation ($\times0.4$), a 40 ms delay and reversal of channel order. Columns show, in order, perturbed strain, logarithmic amplitude ratio, circular phase difference and Riesz orientation difference, relative to the unmodified field. Solid and dotted rectangles delimit the perturbation support and untapered interior. The attributes reveal distinct responses.}
      \label{fig:perturbation_identification}
\end{figure*}

Beyond characterising individual arrivals, the monogenic representation provides a way to compare DAS wavefields through complementary aspects of their structure. Figure~\ref{fig:perturbation_identification}
illustrates this using three localised perturbations of the T2 train: amplitude attenuation, a temporal delay and reversal of channel order. Attenuation produces a nearly uniform negative logarithmic amplitude ratio within the untapered region, close to the expected $\ln(0.4)$, while phase and orientation remain largely unchanged. In comparison, the temporal delay produces a clear phase difference, with comparatively small orientation changes in the interior. Finally, reversing the channel order changes the inclination of the oscillations, producing a strong orientation difference. In essence, the contrasting responses of the attributes help distinguish a change in signal strength from a timing discrepancy or a change in propagation pattern, providing a tool for interpreting information that is otherwise entangled in the recorded wavefield.

This example also suggests a potential application in the evaluation of lossy DAS data compression, a current challenge in DAS deployments \citep{segui2026daspack}. When evaluating the reconstructed field, a small mean-square error may not imply that the oscillatory features used in subsequent analysis have been preserved \citep{segui2026towardspreserving}. Comparing monogenic attributes before and after compression could reveal whether amplitude envelopes, phase continuity and local moveout remain consistent. As such, the representation could supplement sample-wise error measures with criteria related to the intended use of the data, including phase association and apparent-slowness estimation. Further, to achieve very low bit-rate settings, information could be traded-off between quantities, for example spending bits for phase or slowness fidelity at the expense of amplitude fidelity \citep{segui2026geometric}. 

The same reasoning suggests a potential role in constructing misfit functions for inversion. Amplitude, phase and orientation residuals could be combined with separate weights, allowing the comparison to emphasise different aspects of the recorded wavefield. This could extend the single-trace time-frequency misfit framework by \citet{kristekova2009timefrequency}, which is based on Hilbert transform analysis, to the two-dimensional DAS wavefield. For regions
dominated by a single coherent arrival, this could improve the selectivity of recently
proposed spatio--temporal DAS misfits
\citep{ShimonyLellouch2026}, by also using directional coherency to emphasise contributions with consistent local propagation.

\subsection{Further applications}

We think that the proposed methodology has great potential beyond the applications presented here. Monogenic phase and orientation could assist the association of body-wave picks across channels, complementing approaches based on polynomial fits \citep{latorre2025kvp}. The locally estimated moveout could help follow arrivals through changes in cable direction. Directional coherency could also assist the analysis of weak but spatially consistent signals, such as small earthquakes or secondary seismic arrivals.

The monogenic attributes could further provide interpretable features for downstream tasks, including channel-quality evaluation \citep{rodriguez2025cqidas} and training machine-learning models. Their
complementary information could help distinguish changes in signal strength from changes in phase or directional structure. Another possibility is tracking moving sources, including road vehicles and vessels, where the evolution of local moveout could complement existing detection and tracking methods.

\section{Conclusions}

We have presented a methodology for analysing DAS recordings through the monogenic signal, combining the monogenic amplitude, phase and orientation with directional coherency and apparent-slowness estimates. These complementary attributes provide a locally interpretable description of the recorded wavefield, allowing signal strength, oscillatory structure and propagation to be examined in conjunction.

Through submarine \textit{T}-wave recordings, we demonstrated how these attributes support coherent-signal enhancement and directional analysis. Their combination revealed weaker arrivals beneath more energetic wave trains and provided spatially consistent slowness estimates for
investigating incoming directions. We also explored how the representation can distinguish differences between records that would otherwise be combined in a single waveform error.

The proposed methodology is straightforward to incorporate into existing workflows and offers considerable potential to enhance analysis pipelines across a wide range of applications, beyond the examples discussed in this paper. We expect that monogenic analysis could become a meaningful part of the toolkit for turning DAS data into interpretable information.

\begin{acknowledgments}
    Aleix Seguí received the support of a fellowship from the ``la Caixa" Foundation (ID 100010434), under fellowship code LCF/BQ/PFA25/110000. This work was realised during an internship at AP Sensing GmbH. Aleix Seguí thanks Sergi Ventosa and Georgios Kalimeris for helpful discussions.

    Funding for this project was provided through the ‘Severo Ochoa’ grant for excellence CEX2024-001494-S funded by AEI 10.13039/501100011033 and the MCIN/AEI under projects FERMAT (ref. PID2024-162301OB-C21) and MARLIN (CPP2025-012686). We thank Telxius Cable España, S. L. for their support in providing access to the fibre-optic cables at Estepona-Tetouan. Aragón Photonics provided the HDAS interrogator used in the experiment. This work is a contribution of the Barcelona CSI (2025 SGR 00834), a research group recognized by the Generalitat de Catalunya.

    OpenAI's ChatGPT (GPT-6 Astra) was used for language editing, and OpenAI's Codex for script development assistance. All AI-assisted outputs were reviewed and validated by the authors, who retain full responsibility for the manuscript and results.
\end{acknowledgments}

\begin{dataavailability}

Bathymetric data are from the EMODnet Digital Bathymetry DTM 2024 \citep{EMODnetBathymetry2024}. The maps were created using the Generic Mapping Tools (GMT) software \citep{Wessel2019GMT6}. The IGN catalogue is accessible online at \url{https://www.ign.es/web/ign/portal/sis-catalogo-terremotos}. The DAS data from the Estepona-Tetouan \citep{ugalde2023safe} deployment is available from the authors on reasonable request. The scripts implementing the methods described in this manuscript are available at \url{https://github.com/asleix/RieszLab} and data samples at \citet{ugalde2026monogenicdataset}. 

\end{dataavailability}

\bibliographystyle{gji}
\bibliography{references.bib}

\appendix

\section{Analytical derivations}

\subsection{Properties of the Riesz transform}
\label{sec:properties_riesz}

In this section, we briefly establish the main properties of the Riesz transform and discuss their interpretation for spatio--temporal fields. We draw inspiration from \citet{Grafakos2014} for our derivations.

We work with finite-energy functions
$f\in L^2(\mathbb{R}^2)$, where
$\mathbf{x}=(x_1,x_2)$ denotes the two-dimensional coordinate
vector. We use the inner product and associated norm
\begin{equation}
    \langle f,h\rangle
    =
    \int_{\mathbb{R}^2}
    f(\mathbf{x})\overline{h(\mathbf{x})}\,d\mathbf{x},
    \qquad
    \|f\|_2^2=\langle f,f\rangle,
\end{equation}
with the squared norm of a vector-valued field defined as the sum of
the squared norms of its components.

Denoting by $F=\mathcal{F}\{f\}$ the Fourier transform of $f$,
we adopt the convention
\begin{equation}
    F(\boldsymbol{\omega})
    =
    \int_{\mathbb{R}^2}
    f(\mathbf{x})e^{-i\boldsymbol{\omega}\cdot\mathbf{x}}
    \,d\mathbf{x}.
\end{equation}
The Riesz transform
$\mathcal{R}=(\mathcal{R}_1,\mathcal{R}_2)^{\mathsf T}$
is then defined through its Fourier multipliers,
\begin{equation}
    \mathcal{F}\{\mathcal{R}_j f\}(\boldsymbol{\omega})
    =
    m_j(\boldsymbol{\omega})F(\boldsymbol{\omega}),
    \quad
    m_j(\boldsymbol{\omega})
    =
    -i\frac{\omega_j}{\|\boldsymbol{\omega}\|},
    \label{eq:app_riesz_multiplier}
\end{equation}
for $j=1,2$, with the multipliers set to zero at
$\boldsymbol{\omega}=\mathbf{0}$. For convenience, we drop the curly brackets from the Riesz operator notation when there is no ambiguity. 

We study the properties of the transform primarily through
these multipliers, extending familiar properties of the
Fourier transform. We also make use of Parseval's theorem,
which relates the inner products in the spatial and Fourier
domains as
\begin{equation}
    \langle f,h\rangle
    =
    \frac{1}{(2\pi)^2}
    \int_{\mathbb{R}^2}
    F(\boldsymbol{\omega})
    \overline{H(\boldsymbol{\omega})}
    \,d\boldsymbol{\omega}.
    \label{eq:app_parseval}
\end{equation}

\textbf{Linearity.}
Because the Fourier transform and multiplication by
$m_j$ are linear operations, the Riesz transform satisfies
\begin{equation}
    \mathcal{R}\{\alpha f+\beta h\}
    =
    \alpha\mathcal{R}\{f\}
    +
    \beta\mathcal{R}\{h\}.
\end{equation}
For a spatio--temporal wavefield, this means that the Riesz
response of a superposition of waves is the sum of their
individual responses. However, the amplitude, phase and
orientation obtained from the resulting monogenic signal
are not themselves additive. Thus, when several wave trains
overlap, their contributions may become mixed in the
estimated attributes.

\textbf{Composition.}
Applying two Riesz components successively corresponds to
multiplying their Fourier responses. Therefore,
\begin{equation}
    \mathcal{F}\{\mathcal{R}_j\mathcal{R}_k f\}
    =
    -\frac{\omega_j\omega_k}
    {\|\boldsymbol{\omega}\|^2}
    F(\boldsymbol{\omega}).
\end{equation}
In particular, summing the squared components gives
\begin{equation}
    \mathcal{R}_1^2+\mathcal{R}_2^2
    =
    -\mathrm{Id}
    \label{eq:app_riesz_composition}
\end{equation}
where $\mathrm{Id}$ denotes the identity operator on
$L^2(\mathbb{R}^2)$. This identity is the two-dimensional
counterpart of $\mathcal{H}^2=-\mathrm{Id}$ for the
Hilbert transform. This property is intuitive, because performing two consecutive phase-shifts of $90^\circ$ corresponds to phase-shift of $180^\circ$, that is, a sign inversion.   

\textbf{Adjoint.}
The adjoint of an operator $\mathcal{T}$ is defined through
the relation
$\langle\mathcal{T}f,h\rangle
=\langle f,\mathcal{T}^{\dagger}h\rangle$.
As the Riesz multipliers are purely imaginary,
$\overline{m_j}=-m_j$, and Parseval's theorem gives
\begin{equation}
\begin{aligned}
    \langle\mathcal{R}_j \{f\},h\rangle
    &=
    \frac{1}{(2\pi)^2}
    \int_{\mathbb{R}^2}
    m_j F\overline{H}\,d\boldsymbol{\omega}\\
    &=
    -\frac{1}{(2\pi)^2}
    \int_{\mathbb{R}^2}
    F\overline{m_j H}\,d\boldsymbol{\omega} =
    -\langle f,\mathcal{R}_j\{h\}\rangle.
\end{aligned}
\end{equation}
That is, each Riesz component is anti-Hermitian,
$\mathcal{R}_j^{\dagger}=-\mathcal{R}_j$. For the vector-valued Riesz transform, the adjoint is
defined through the sum of the component-wise inner products,
\begin{equation}
\begin{aligned}
    \langle\mathcal{R}f,\mathbf{u}\rangle
    &=
    \langle\mathcal{R}_1 f,u_1\rangle + \langle\mathcal{R}_2 f,u_2\rangle\\
    &=
    \left\langle f,
    -\mathcal{R}_1 u_1 -\mathcal{R}_2 u_2
    \right\rangle.
\end{aligned}
\end{equation}
Hence, its adjoint maps a two-component field
$\mathbf{u}=(u_1,u_2)^{\mathsf T}$
back to a scalar field as $\mathcal{R}^{\dagger}\mathbf{u} = -\mathcal{R}_1u_1-\mathcal{R}_2u_2.$ Combining this with the composition identity yields the useful relation
\begin{equation}
    \mathcal{R}^{\dagger}\mathcal{R}
    =
    \mathcal{R}_1^{\dagger}\mathcal{R}_1
    +\mathcal{R}_2^{\dagger}\mathcal{R}_2
    =
    -\mathcal{R}_1^2-\mathcal{R}_2^2
    =
    \mathrm{Id}.
    \label{eq:app_riesz_adjoint_identity}
\end{equation}
This will later be helpful for the
reconstruction of strain from the enhanced monogenic signal.

\textbf{Isometry.}
From the preceding identity, we obtain
\begin{equation}
    \|\mathcal{R}f\|_2^2
    =
    \langle\mathcal{R}f,\mathcal{R}f\rangle
    =
    \langle f,\mathcal{R}^{\dagger}\mathcal{R}f\rangle
    =
    \|f\|_2^2.
\label{eq:app_riesz_isometry}
\end{equation}
Thus, the Riesz components together preserve the
$L^2$ energy of the original field, redistributing the signal energy
between each component.

\textbf{Translation invariance.}
Consider a translation
$f_{\mathbf{b}}(\mathbf{x})=f(\mathbf{x}-\mathbf{b})$.
Its Fourier transform is
$F_{\mathbf{b}}(\boldsymbol{\omega})
=e^{-i\boldsymbol{\omega}\cdot\mathbf{b}}
F(\boldsymbol{\omega})$.
Because this exponential factor commutes with the Riesz
multipliers, we obtain
\begin{equation}
    \mathcal{R}\{f_{\mathbf{b}}\}(\mathbf{x})
    =
    \mathcal{R}\{f\}(\mathbf{x}-\mathbf{b}).
\end{equation}
Therefore, translating a wavefield in space or time
produces the corresponding translation of its Riesz
components.

\textbf{Scale invariance.}
Consider a uniform dilation
$f_a(\mathbf{x})=f(\mathbf{x}/a)$, with $a>0$.
Since its Fourier transform satisfies
$F_a(\boldsymbol{\omega})=a^2F(a\boldsymbol{\omega})$,
and the Riesz multipliers 
$m_j(a\boldsymbol{\omega})=m_j(\boldsymbol{\omega})$,
we obtain
\begin{equation}
    \mathcal{R}\{f_a\}(\mathbf{x})
    =
    \mathcal{R}\{f\}(\mathbf{x}/a).
\end{equation}
Thus, a uniform dilation of the two wavefield components produces the
same dilation of the Riesz response. Notice, however, that this property does not extend to independent scaling of the two coordinates. Indeed, changing
only the spatial or temporal scale modifies the relative
frequency components in $m_j$, and hence
the Riesz orientation. We highlighted this property
in spatio--temporal DAS fields, where the relative
scaling of distance and time determines the velocity metric
of the analysis (Sections \ref{sec:monogenic_das}~and~\ref{sec:vel_metric}).

\textbf{Rotation covariance.}
Finally, consider a rotation represented by an orthogonal
matrix $\mathbf{Q}$, and define
$f_{\mathbf{Q}}(\mathbf{x})
=f(\mathbf{Q}^{\mathsf T}\mathbf{x})$.
Its Fourier representation satisfies
$F_{\mathbf{Q}}(\boldsymbol{\omega})
=F(\mathbf{Q}^{\mathsf T}\boldsymbol{\omega})$.
Because rotations preserve the Euclidean norm, the vector
of Riesz multipliers
$\mathbf{m}(\boldsymbol{\omega})
=-i\boldsymbol{\omega}/\|\boldsymbol{\omega}\|$
satisfies
$\mathbf{m}(\boldsymbol{\omega})
=\mathbf{Q}\mathbf{m}(\mathbf{Q}^{\mathsf T}\boldsymbol{\omega})$, giving us
\begin{equation}
    \mathcal{R}\{f_{\mathbf{Q}}\}(\mathbf{x})
    =
    \mathbf{Q}\,
    \mathcal{R}\{f\}(\mathbf{Q}^{\mathsf T}\mathbf{x}).
\end{equation}
Hence, when a field is rotated, its Riesz vector components
rotate with the field: the magnitude of the Riesz
vector is preserved, while its orientation follows
the rotation of the field.

\subsection{Derivation of the enhanced field}
\label{app:monogenic_enhancement}

In the main text, we introduced signal enhancement through the use of directional coherency as a spatio--temporal weight.  Now, we provide details on how to obtain the enhanced field from the higher-dimensional monogenic representation, using the properties
of the Riesz transform established above.

Let $f$ denote the original real-valued strain field.
Our proposed enhancement consists of multiplying the scalar field
and its two Riesz components by a spatially varying
weight $\mathcal{X}(\mathbf{x})$, representing the
directional coherency or its combination with an
orientation-selective weight. We assume that
$\mathcal{X}$ is real and bounded.

Because the Riesz transform is nonlocal, multiplication
by a spatially varying weight does not generally commute
with the transform:
$\mathcal{R}_j \{ \mathcal{X}f \}
\neq \mathcal{X}\mathcal{R}_j \{ f \}$.
Consequently, the enhanced components need not correspond
to the monogenic representation of a single strain field.
For this reason, we seek a field $u$ whose Riesz components
best approximate the enhanced ones.

A natural choice is to find the closest consistent
monogenic representation in the least-squares sense.
Intuitively, this can be understood as projecting the
weighted components onto the subspace of valid monogenic
representations, that is, those arising from a single
strain field. Based on this principle, we define the
least-squares functional
\begin{equation}
    J(u)
    =
    \|u-\mathcal{X}f\|_2^2
    +
    \|\mathcal{R}\{u\}
    -\mathcal{X}\mathcal{R}\{f\}\|_2^2,
    \label{eq:app_reconstruction_functional}
\end{equation}
which measures the total squared discrepancy between
the monogenic components of a candidate field $u$
and their weighted counterparts. We seek its minimiser over
real-valued finite-energy fields,
\begin{equation}
    f_{\star}
=\operatorname{arg\,min}_{u\in L^2(\mathbb{R}^2;\mathbb{R})}J(u),
\end{equation}
where $f_{\star}$ is the enhanced strain field.

We employ a technique from calculus of variations to
determine the function that minimises this functional. It consists in first identifying a stationary point of $J(u)$ through its
first variation, and then verifying that it is the
unique global minimum \citep[see, e.g., ][]{luenberger1969optimization}. We choose this approach because it naturally extends to more general objectives, including additional constraints or regularisation.

To this end, we consider a perturbation
$u+\varepsilon h$, where $h$ is an arbitrary
real-valued finite-energy field and $\varepsilon$
is a real scalar. By linearity of the Riesz transform,
the perturbation introduces the terms $\varepsilon h$
and $\varepsilon\mathcal{R}\{h\}$ into the two residuals
of the functional. Expanding their squared norms gives
\begin{equation}
\begin{split}
    J(u+\varepsilon h)
    &=
    J(u)
    +2\varepsilon\operatorname{Re}
    \Bigl[
        \langle h,u-\mathcal{X}f\rangle\\
    &\qquad+
        \langle\mathcal{R}\{h\},
        \mathcal{R}\{u\}-\mathcal{X}\mathcal{R}\{f\}\rangle
    \Bigr]\\
    &\quad+
    \varepsilon^2
    \left(
        \|h\|_2^2+\|\mathcal{R}\{h\}\|_2^2
    \right).
\end{split}
\label{eq:app_reconstruction_expansion}
\end{equation}
The first variation is the derivative of this expression
with respect to $\varepsilon$, evaluated at
$\varepsilon=0$. It describes the first-order change
in the functional produced by an infinitesimal
perturbation of $u$ along the direction $h$.
Using the definition of the adjoint established above,
we obtain
\begin{equation}
\begin{split}
    \delta J(u;h)
    &=
    \left.
    \frac{d}{d\varepsilon}
    J(u+\varepsilon h)
    \right|_{\varepsilon=0}\\
    &=
    2\operatorname{Re}
    \left\langle
        h,\,
        u-\mathcal{X}f
        +\mathcal{R}^{\dagger}
        \bigl(
            \mathcal{R}\{u\}
            -\mathcal{X}\mathcal{R}\{f\}
        \bigr)
    \right\rangle.
\end{split}
\label{eq:app_reconstruction_variation}
\end{equation}

At a minimum, the first variation must vanish for
every admissible perturbation $h$. Indeed, since
$\varepsilon$ may take either sign, a nonzero derivative
would imply that a sufficiently small perturbation
in one direction decreases the functional.
Furthermore, because $h$ is arbitrary, we can choose
it equal to the field appearing in the inner product. Hence, as the first variation must vanish at stationary points, this requires
\begin{equation}
    u-\mathcal{X}f
    +\mathcal{R}^{\dagger}
    \bigl(
        \mathcal{R}\{u\}
        -\mathcal{X}\mathcal{R}\{f\}
    \bigr)
    =0.
\label{eq:app_reconstruction_stationarity}
\end{equation}
Using the identities in
Equation~\ref{eq:app_riesz_adjoint_identity},
the stationary condition is simplified and we can solve for $u$ to obtain enhanced strain field
\begin{equation}
    f_{\star}
    =
    \frac{1}{2}
        \mathcal{X}f
        +
        \frac{1}{2}
        \mathcal{R}^\dagger
        \{
            \mathcal{X}\mathcal{R}\{f\}
        \}.
\label{eq:app_monogenic_reconstruction}
\end{equation}
Thus, we find that the enhanced field combines the directly weighted strain with the weighted Riesz components mapped back
to a scalar field through the adjoint. In particular,
when $\mathcal{X}=1$, the composition identity recovers
the original strain.

It remains to verify that this stationary point
is indeed the global minimum. Expanding the functional
around $f_{\star}$, the first-order term vanishes
by stationarity. Using the isometry of the vector
Riesz transform
(Equation~\ref{eq:app_riesz_isometry}),
we obtain, for any admissible field $u$,
\begin{equation}
\begin{split}
    J(u)-J(f_{\star})
    &=
    \|u-f_{\star}\|_2^2
    +\|\mathcal{R}(u-f_{\star})\|_2^2\\
    &=
    2\|u-f_{\star}\|_2^2
    \geq 0.
\end{split}
\label{eq:app_reconstruction_optimality}
\end{equation}
Equality holds only when $u=f_{\star}$, proving that the enhanced strain field is the unique global minimiser.

\label{lastpage}

\end{document}